\documentclass[onefignum,onetabnum]{siamart171218}

\usepackage{lipsum}
\usepackage{amsfonts}
\usepackage{graphicx}
\usepackage{epstopdf}
\usepackage{algorithmic}
\ifpdf
  \DeclareGraphicsExtensions{.eps,.pdf,.png,.jpg}
\else
  \DeclareGraphicsExtensions{.eps}
\fi

\graphicspath{{./figures/}}

\newsiamremark{remark}{Remark}
\newsiamremark{hypothesis}{Hypothesis}
\crefname{hypothesis}{Hypothesis}{Hypotheses}
\newsiamthm{claim}{Claim}

\title{Accelerating wave-propagation algorithms with adaptive mesh refinement using the Graphics Processing Unit (GPU)}

\author{
    Xinsheng Qin\thanks{Department of Civil and Environmental Engineering, 
    University of Washington, Seattle, WA 
    (\email{xsqin@uw.edu}).}
\and 
    Randall LeVeque\thanks{Department of Applied Mathematics, 
    University of Washington, Seattle, WA 
    (\email{rjl@uw.edu}).}
\and 
    Michael Motley\thanks{Department of Civil and Environmental Engineering, 
    University of Washington, Seattle, WA 
    (\email{mrmotley@uw.edu}).}
}

\usepackage{amsopn}

\usepackage{amsfonts,amssymb,amsmath}
\usepackage{dsfont}
\usepackage{graphicx}
\usepackage{hyperref}
\usepackage[caption=false,font=footnotesize]{subfig}

\newcommand{\ignore}[1]{}

\newenvironment{mat}{\left[ \begin{array}{ccccccccccccc}}{\end{array}\right]}
\newenvironment{rmat}{\left[ \begin{array}{rrrrrrrrrrrrr}}{\end{array}\right]}
\newenvironment{lmat}{\left[ \begin{array}{lllllllllllll}}{\end{array}\right]}
\newcommand\bcm{\begin{mat}}
\newcommand\ecm{\end{mat}}
\newcommand\brm{\begin{rmat}}
\newcommand\erm{\end{rmat}}
\newcommand\blm{\begin{lmat}}
\newcommand\elm{\end{lmat}}

\newenvironment{pwdef}{\left\{ \begin{array}{ll}}{\end{array}\right.}
\newcommand\bpwdef{\begin{pwdef}}
\newcommand\epwdef{\end{pwdef}}

\newcommand\bc{\begin{center}}
\newcommand\ec{\end{center}}
\newcommand\bi{\begin{itemize}}
\newcommand\ei{\end{itemize}}
\newcommand\be{\begin{enumerate}}
\newcommand\ee{\end{enumerate}}

\newcommand\bsplit{\begin{split}}
\newcommand\esplit{\end{split}}

\newcommand{\eq}{\begin{equation}}
\newcommand{\en}{\end{equation}}
\newcommand{\eqm}{\begin{eqnarray}}
\newcommand{\enm}{\end{eqnarray}}
\newcommand{\eqmno}{\begin{eqnarray*}}
\newcommand{\enmno}{\end{eqnarray*}}
\newcommand{\eqml}[1]{\eql{#1}\begin{array}{rcl}}
\newcommand{\enml}{\end{array}\en}
\newcommand{\eql}{\begin{equation}\label}
\newcommand{\eqsub}[1]{\begin{subequations}\label{#1}\eqm }
\newcommand{\ensub}{\enm\end{subequations}}

\newcommand\reals{{{\rm l} \kern -.15em {\rm R} }}

\newcommand\Dx{\Delta x}

\newcommand\qij{\Q_{ij}}

\newcommand\qijnp{\Q_{ij}^{n+1}}

\newcommand\apdq{{\cal A}^+\Delta \Q}
\newcommand\amdq{{\cal A}^-\Delta \Q}
\newcommand\bpdq{{\cal B}^+\Delta \Q}
\newcommand\bmdq{{\cal B}^-\Delta \Q}

\newcounter{equationgroup}
\newcommand\Q{Q}

\newcommand{\tFimhjn}{\tilde F_{i-1/2,j}^n}
\newcommand{\tFiphjn}{\tilde F_{i+1/2,j}^n}
\newcommand{\tGijphn}{\tilde G_{i,j+1/2}^n}
\newcommand{\tGijmhn}{\tilde G_{i,j-1/2}^n}

\newcommand{\apdqjmh}{{\cal A}^+\Delta \Q_{j-1/2}}

\newcommand{\apdqmphb}{{\cal A}^+\Delta \Q_{m+1/2,b}}
\newcommand{\apdqmphbp}{{\cal A}^+\Delta \Q_{m+1/2,b+1}}

\newcommand\apdqimhj{\apdq_{i-1/2,j}}

\newcommand\amdqiphj{\amdq_{i+1/2,j}}

\newcommand\bpdqijmh{\bpdq_{i,j-1/2}}

\newcommand\bmdqijph{\bmdq_{i,j+1/2}}

\newcommand{\ico}[1]{#1 \kern -1ex \raisebox{1.1ex}{$\circ$}}
\newcommand{\icos}[2]{#1 \kern -1.1ex \raisebox{1.1ex}[.5em]{$\circ$}^{#2}}

\begin{document}

\maketitle

\begin{abstract}
    Clawpack is a library for solving nonlinear hyperbolic partial differential equations using high-resolution finite volume methods based on Riemann solvers and limiters. 
    It supports Adaptive Mesh Refinement (AMR), which is essential in solving multi-scale problems. 
    Recently, we added capabilities to accelerate the code by using the Graphics Process Unit (GPU) \footnote{The code and benchmark used in this study can be found at \url{https://github.com/xinshengqin/amrclaw/tree/gpu_amr_paper_benchmark_tag/examples/GPU/acoustics_2d_radial}.}.
    Routines that manage CPU and GPU AMR data and facilitate the execution of GPU kernels are added. 
    Customized and CPU thread-safe memory managers are designed to manage GPU and CPU memory pools, which is essential in eliminating the overhead of memory allocation and de-allocation. 
    A global reduction is conducted every time step for dynamically adjusting the time step based on Courant number restrictions. 
    Some small GPU kernels are merged into bigger kernels, which greatly reduces kernel launching overhead.
    A speed-up between $2$ and $3$ for the total running time is observed in an acoustics benchmark problem.
\end{abstract}



\section{Introduction} \label{sec:introduction}
Many problems exhibit multi-scale behavior. 
The modeling of such problems with discretized Partial Differential Equations (PDEs) requires higher spatial and temporal resolution where the solution has localized large errors. 
Adaptive Mesh Refinement (AMR) is a popular approach for tracking features much smaller than the overall scale of the problem and adjusting the computational grid adaptively.

There are three major variants of AMR implementations.
The first one is often referred to as patch-based or structured AMR \cite{berger1989local}.
It allows rectangular grid patches of arbitrary size and any integer refinement ratios between two level of grid patches.
This approach has been implemented by many AMR packages \cite{beckingsale2014parallel,bryan2014enzo}, including the Clawpack software that is the focus of this work \cite{clawpack}.
The second type is cell-based AMR which refines individual cells and often uses octree data structure to store the grid patch information.
The last type is a combination of the first two, and uses an octree to store the grid patch hierarchy information and requires all grid patches to be of the same size (e.g. 8 by 8 in 2D case) \cite{burstedde2014forestclaw,burstedde2010extreme,Calhoun2017,fryxell2000flash,schive2010gamer}.

Recently, the use of Graphics Processing Units (GPUs) has been shown to be very successful in accelerating scientific computing in many fields.
When they are used to accelerate PDE solvers with AMR grids, a big challenge comes from the complexity of the AMR algorithm and data structure.

In the last decade, many groups have developed packages that can solve PDEs with AMR grids on the GPU.
For example, Wang et al. \cite{wang2010adaptive} implemented an inviscid fluid solver with patch-based AMR on the GPU, which runs $10$ times faster on a Quadro FX 5600 GPU than its CPU version running on a single 3 GHz CPU core, and
Schive et al. \cite{schive2010gamer} described the development of an AMR package called GAMER on the GPU. 
The GPU implementation increased the code performance by nearly one order of magnitude in a benchmark of purely baryonic cosmological simulations, using 1-32 Tesla T10 GPUs versus the same number of 4-core Intel Xeon CPUs.
Recently, they reported the second version of the code \cite{schive2017gamer}, which has better load balancing, includes more fluid solvers, and scales well to thousands of GPUs.
Bryan et al. \cite{bryan2014enzo} introduced ENZO: a patch-based AMR library for astrophysics.
The results of a weak scaling test of driven MHD turbulence on a uniform mesh showed a speed-up by a factor of $\sim 5$ when running on 1--8 NVIDIA K20 GPUs compared to the same number of 16-core AMD Opteron CPUs.

The packages above all share the same design strategy that all cell data is transferred back and forth between the CPU and GPU memory, since most of the AMR algorithm except advancing the solution on each grid patch are still performed by the CPUs.
Another strategy is to keep all cell data in the GPU memory during the entire simulation, which is often referred to as resident implementation of AMR on the GPU.
Beckingsale et al. \cite{beckingsale2015resident,beckingsale2014parallel} implemented a resident patch-based AMR library on the GPUs,
and achieved about two times speed-up when running on a single NVIDIA K20x GPU versus a dual-socket Intel Xeon at 2.6GHz (16 CPU cores in total). 
The code is also shown to scale well to thousands of nodes in a weak scaling test.

Clawpack \cite{clawpack,mandli2016clawpack} is a software package designed to solve linear and nonlinear hyperbolic PDEs using high-resolution finite volume methods based on Riemann solvers and limiters. 
It has been actively developed as an open source project for over 20 years.
The underlying solvers are based on the wave propagation algorithms described in \cite{leveque2002finite}, and it incorporates patch-based AMR using the algorithms described in \cite{BergerLeVeque1998}.
Recently the ManyClaw  project \cite{terrel2013manyclaw} has explored the exploitation of intra-node parallelism in hyperbolic PDE solvers via Clawpack.

A CUDA implementation of the single-grid version of Clawpack has previously been developed in the CUDACLAW project \cite{ohannessian2018cudaclaw}, but that work did not attempt to handle the AMR portions of Clawpack.  This paper describes a strategy and implementation for doing so, and is structured as follows.
\Cref{sec:wave_and_amr} gives an overview of the wave propagation algorithm and how it is combined with the AMR algorithm in Clawpack.
\Cref{sec:design} explains how the code was modified for running on the GPU, extra routines that were added, how a new GPU kernel for advancing the solution of a particular hyperbolic problem can be written, and some key strategies for better performance.
\Cref{sec:benchmark} discuss two AMR parameters that affect the performance of the code and evaluate its performance on 2 different GPUs.

\section{Wave-Propagation algorithm on Adaptive Mesh} \label{sec:wave_and_amr}
\subsection{Hyperbolic PDEs and Wave-Propagation Algorithm} \label{sec:wave_propatation}
Hyperbolic PDEs arise in the modeling of many scientific and engineering problems. 
Since the solution of nonlinear problems usually involves shock formation, discontinuities in the solution can arise even if the initial data are smooth.  As a result, expensive high-resolution shock capturing methods coupled with adaptive refinement are often used to numerically solve hyperbolic PDEs, and these methods also work well for linear problems, particularly in heterogeneous media with discontinuous coefficients, for example.
In this study, a wave-propagation form of Godunov's method is used, details of which can be found in \cite{BergerLeVeque1998,leveque2002finite}. 
A brief overview is given below, using the same notation as in these references.

For simplicity, we introduce a two dimensional system of hyperbolic equations in conservative form:
\begin{equation}
    q_t + f(q)_x + g(q)_y= 0
    \label{eq:hyperbolic}
\end{equation}
where $q(x,y,t) \in \mathds{R}^m$ is a vector with $m$ components representing the unknowns, and $f(q)$ and $g(q)$ are flux functions in the $x$- and $y$-directions. 
The subscripts represent partial derivatives with respect to $x$- and $y$ respectively.
In a finite volume method, the solution at time $t_n$ is viewed as an approximation to the average over the grid cell:
\begin{equation}
    \qij = \frac{1}{\Delta x \Delta y} \int_{y_{j-1/2}}^{y_{j+1/2}} \int_{x_{i-1/2}}^{x_{i+1/2}} q(x,y,t^n)\, dx\, dy.
\end{equation}
Riemann problems need to be solved between cells to get a set of traveling waves at each cell edge, which are then used in updating cell values.
The solution can be updated with fluctuations (waves) that propagate across cell edges and 2nd-order correction terms in flux form:
{\footnotesize
\begin{align}
\begin{split}
    \qijnp = & \qij - \frac{\Delta t}{\Delta x} \left( \apdqimhj + \amdqiphj \right) - \frac{\Delta t}{\Delta y} \left( \bpdqijmh + \bmdqijph \right) \\ 
             & - \frac{\Delta t}{\Delta x} \left( \tFiphjn - \tFimhjn \right) - \frac{\Delta t}{\Delta y} \left( \tGijphn - \tGijmhn \right)   
    \label{eq:wave_form_2d_2nd_order}
\end{split}
\end{align}
}
where $\apdqimhj$, $\amdqiphj$, $\bpdqijmh$ and $\bmdqijph$ represent the total waves from the four edges of cell $(i,j)$ that travel into the cell, and $\tFiphjn$, $\tFimhjn$, $\tGijphn$ and $\tGijmhn$ are 2nd-order correction terms also from four edges of cell $(i,j)$ but in flux form. 
For instance, $\apdqimhj$ represents the right-going wave from the Riemann problem solved at edge $\left( i-1/2, j \right)$. 
The 2nd-order correction terms include both modification to the waves arising from representing cell values with piecewise linear function and transverse waves from splitting the original waves in the transverse direction. 
Details of these can be found in \cite{leveque2002finite}.

\subsection{Adaptive Mesh Refinement (AMR)} \label{sec:amr}
Block-structured adaptive mesh refinement (AMR) is a popular approach for tracking features much smaller than overall scale of the problem and adjusting the computational grid adaptively.
The algorithm is fully described in \cite{berger1989local,BergerLeVeque1998} and is only briefly summarized here.

In this approach, a collection of rectangular grid patches are used to store the solution. 
All grid patches are uniform and have different levels of resolution.
The coarsest grid patches cover the entire domain. 
The 2nd coarsest grid patches cover a subset of the domain and are nested in the coarsest grid patches.
Finer grid patches are nested recursively and cover smaller subsets of the domain.
Typically, if level $L+1$ grid patches are $R_{L}$ times finer than level $L$ grid patches in $x$- and $y$-directions, the time step is also smaller by the same factor (hereafter $R_L$ represents the refinement ratio in the $x$- and $y$-directions between level $L$ and level $L+1$ grid patches).

Every $K$ time steps on a particular grid level, all finer level grid patches will be regenerated since features of the flow might have moved.
Cells are flagged for refinement using some criterion (e.g., where the gradient or an estimate of the one-step error are above some specified tolerance). 
The flagged cells are then clustered into new rectangular grid patches,
which usually include some cells not flagged as well, using an algorithm of \cite{BergerRigoutsos}.

The wave-propagation form of Godunov's method described in the previous section can be then used to advance the solution on the grid hierarchy.
Grid patches are advanced level by level with the coarser level advanced first.
Level $1$ grid patches are advanced by one time step first. 
Then level $2$ grid patches are advanced by $R_1$ steps (after which the solution on level $2$ grid patches is at the same time as the level $1$ grid patches) before level $1$ grid patches are advanced again.
This is applied to all levels of grid patches recursively.
For instance, if there is a level $3$, then these grid patches are advanced by $R_2$ steps between each two time steps of level $2$ grid patches.

When level $L$ grid patches catch up with coarser level $L-1$ grid patches in time, cell values in level $L$ grid patches are used to update overlapping level $L-1$ grid patches by simple averaging, since a more accurate solution is available.
This process is referred to as {\em updating} hereafter.
The {\em updating} process between level $L$ and level $L-1$ grid patches makes level $L-1$ grid patches lose global conservation, since some of the cells are simply over-written by new values that were computed using different interface fluxes on the finer grid patches.
A conservation fix must be applied to preserve global conservation at level $L-1$, which is usually achieved by modifying values of some level $L-1$ cells that border the over-written region (coarse and fine grid patches interface).

Boundary conditions for each grid patch can be imposed by extending the grid patch to include a few additional cells on each edge, which are usually called ghost cells.
The ghost cells of each AMR grid patch must be filled with proper values before a time step can be taken for the grid patch. 
The value of each ghost cell is determined by the relative location of the ghost cell with respect to the domain and other grid patches:
\begin{enumerate}
    \item If the ghost cell is outside the computational domain, ghost cell values are determined using physical boundary conditions.
    \item If the ghost cell overlaps other grid patches on the same level, ghost cell values can be copied from the other grid patches.
    \item If the ghost cell interior to the domain does not overlap with any other grid patch at the same level, ghost cell values can be interpolated from underlying coarse grid patches.
\end{enumerate}

\begin{figure}[ht]
\centering
\includegraphics[width=2.5in]{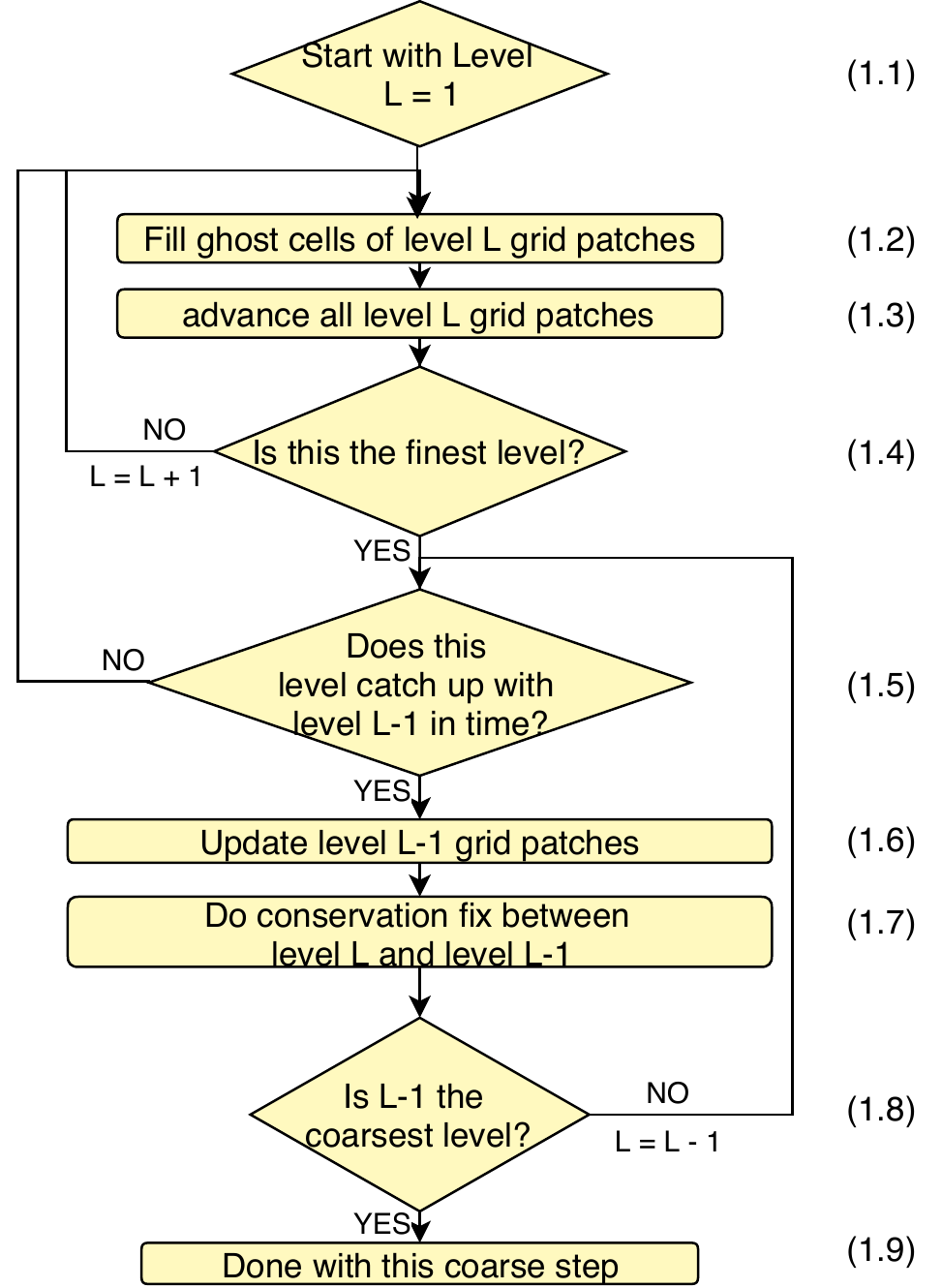}
\caption{A flow chart of AMR algorithm including 1) advancing the solution, 2) {\em updating} and 3) conservation fix. Assume at least two levels of grid patches exist.}
\label{fig:amr_flow_chart}
\end{figure}

The algorithm is summarized in \cref{fig:amr_flow_chart}, where a flow chart shows how the coarsest level grid patches are advanced by one time step.

\subsection{Updating Process and Conservation Fix}

\begin{figure}[!t]
\centering
\includegraphics[width=2.5in]{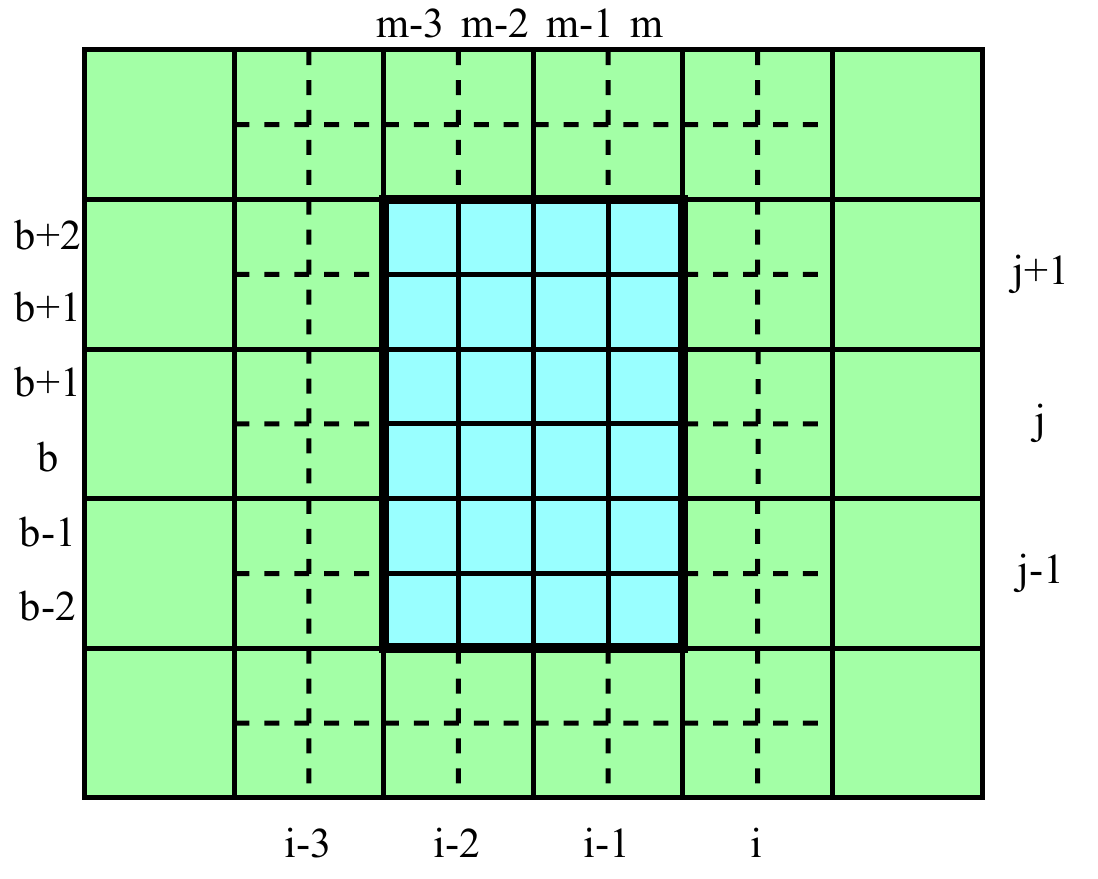}
\caption{A 2D coarse grid patch and a nested fine grid patch. $i$ and $j$ are index for the coarse gird in $x$- and $y$-directions. $m$ and $b$ are index for the fine gird in $x$- and $y$-directions}
\label{fig:2d_grids}
\end{figure}

\Cref{fig:2d_grids} shows a fine grid patch (in cyan) nested in a coarser grid patch (in pale green) with a refinement ratio of $2$ in both $x$- and $y$-directions. 
The ghost cells of the fine grid patch are denoted with dashed lines.
When the fine grid patch catches up with the coarse grid patch in time (answer ``YES'' to Step \ref*{fig:amr_flow_chart}.5 in \cref{fig:amr_flow_chart}), then the solution
values in 6 coarse grid cells covered by the fine grid cells will be over-written by the average values of fine grid cells (Step \ref*{fig:amr_flow_chart}.6 in \cref{fig:amr_flow_chart}). 
For instance,
\begin{equation}
    Q_{i-1,j} := \frac{1}{4}\left( \hat Q_{m-1,b} + \hat Q_{m,b} + \hat Q_{m-1,b+1} \hat Q_{m,b+1} \right)
\end{equation}
where $Q$ is the cell value in the coarse grid patch and $\hat Q$ is the cell value in fine grid patch.
In the conservation fix step, modifications must be added to the 10 coarse cells around the fine grid patch, to preserve global conservation in the coarse grid patch (Step \ref*{fig:amr_flow_chart}.7 in \cref{fig:amr_flow_chart}).
It can be shown that, for example, the modification added to coarse cell $(i,j)$ for a conservation law in \cref{eq:hyperbolic} should be:

\begin{align}
    \qijnp := & \qijnp + C1 + C2 + C3
    \label{eq:c123}
\end{align}

Modification terms are grouped into three categories $C1$, $C2$ and $C3$:
\begin{align}
    \begin{split}
    C1 := & \frac{1}{4} \frac{\Delta t}{\Dx} \left[ \left(  f(\hat{Q}_{m+1,b}^n) - f(Q_{i,j}^n) \right) + \left(  f(\hat{Q}_{m+1,b+1}^n) - f(Q_{i,j}^n) \right) \right] \\
    + & \frac{1}{4} \frac{\Delta t}{\Dx} \left[ \left(  f(\hat{Q}_{m+1,b}^{n+1/2}) - f(Q_{i,j}^n)\right) +  \left(  f(\hat{Q}_{m+1,b+1}^{n+1/2}) - f(Q_{i,j}^n)\right) \right] 
    \end{split}
\end{align}

\begin{subequations}
\begin{align}
    \label{eq:c2_1}
    C2 := & - \frac{1}{4} \frac{\Delta t}{\Dx} \left( \widehat{\apdqmphb ^n} + \widehat{\apdqmphbp ^n} \right)   \\
    \label{eq:c2_2}
              & - \frac{1}{4} \frac{\Delta t}{\Dx} \left( \widehat{\apdqmphb ^{n+1/2}} + \widehat{\apdqmphbp ^{n+1/2}}\right) \\
    \label{eq:c2_3}
              & + \frac{\Delta t}{\Dx}\apdqjmh ^n
\end{align}
\end{subequations}

\begin{subequations}
\begin{align}
    \label{eq:c3_1}
    C3 := & +\frac{1}{4} \frac{\Delta t}{\Dx} \left( \widehat{\tilde F_{m+1/2,b}^{n}} + \widehat{\tilde F_{m+1/2,b+1}^{n}} \right) \\
    \label{eq:c3_2}
    &+\frac{1}{4} \frac{\Delta t}{\Dx} \left(  \widehat{\tilde F_{m+1/2,b}^{n+1/2}} + \widehat{\tilde F_{m+1/2,b+1}^{n+1/2}} \right) \\
    \label{eq:c3_3}
     & - \frac{1}{4} \frac{\Delta t}{\Dx} \tilde F_{i-1/2,j}^n 
\end{align}
\end{subequations}
where $\Delta t$ is the time step size for the coarse grid patch, $\Delta x$ is grid cell size in $x$-direction in the coarse grid patch, 
$Q$ and $\hat{Q}$ are cell values in the coarse and fine grid patch, $\apdq$ and $\widehat{\apdq}$ are right going waves in the coarse and fine grid patch,
and $\tilde F$ and $\widehat{\tilde F}$ are combination of 2nd-order correction terms and transverse waves in flux form for the coarse and fine grid patch. 
The superscript $n+\frac{1}{2}$ represents the time when the fine grid patch is advanced by one time step of size $\frac{\Delta t}{2}$ but has not caught up with the coarse grid patch at time step $n+1$.
A better way to write $C1$ is:
{\small
\begin{subequations}
\begin{align}
    \label{eq:c1_1}
    C1 := & \frac{1}{4} \frac{\Delta t}{\Dx} \left[ \left(  \apdq _b^n + \amdq _b^n \right) + \left(  \apdq _{b+1}^n + \amdq _{b+1}^n \right) \right] \\
    \label{eq:c1_2}
    + & \frac{1}{4} \frac{\Delta t}{\Dx} \left[ \left(  \apdq _b^{n+1/2} + \amdq _b^{n+1/2} \right) +  \left(  \apdq _{b+1}^{n+1/2} + \amdq _{b+1}^{n+1/2} \right) \right] 
\end{align}
\end{subequations}
}
which allows the hyperbolic equations not in conservation form to be handled the same way as those in conservation form.

These modification terms are briefly explained as follows.
Each of the $4$ terms in $C1$ contains waves from solving a Riemann problem between a fine cell and a coarse cell at the time represented by the superscript.
For example, $\apdq _{b+1}^{n+1/2}$ and $\amdq _{b+1}^{n+1/2}$ are right- and left-going waves from solving a Riemann problem with $\hat{Q}_{m+1,b+1}^{n+1/2}$ and $Q_{i,j}^n$ as right and left initial states.
$C2$ includes right-going waves at the coarse-fine interface.
\Cref{eq:c2_1,eq:c2_2} are computed when the fine grid patch is advanced from time $t_n$ and from $t_n+\frac{\Delta t}{2}$, respectively.
\cref{eq:c2_3} are also right-going waves at the coarse-fine interface, but is computed when the coarse grid patch is advanced from time $t_n$.
For example, $\widehat{\apdqmphb ^{n+1/2}}$ is the right-going waves from solving a Riemann problem between cell $(m,b)$ and $(m+1,b)$ when the fine grid patch is advanced from time $t_n + \frac{\Delta t}{2}$ to time $t_{n+1}$. 
We can save these waves at the coarse-fine interface when grids patches are advanced in time instead of recomputing them for later use in the conservation fix process.
$C3$ are second order terms in flux form at the coarse-fine interface from both fine and coarse grid patches. 
The definition of these fluxes are the same as those in \cref{eq:wave_form_2d_2nd_order} which includes both 2nd-order correction terms and transverse waves.
These terms can also be saved when the grid patches are advanced for later use in the conservation fix process.
For example, when cell $(i,j)$ in the coarse grid patch is advanced with \cref{eq:wave_form_2d_2nd_order}, flux $\tilde F_{i-1/2,j}^n$ can be saved and reused as \cref{eq:c3_3} for computing $C3$.

\subsection{Time Step Size}
The time step size $\Delta t$ in \cref{eq:wave_form_2d_2nd_order} must be chosen carefully at each time step if variable time step is used.
The Courant, Friedrichs and Lewy (CFL) condition implies that the time step size must satisfy
\begin{equation}
    \nu \equiv \left| \frac{s \Delta t}{\Delta x} \right| \leq 1
\end{equation}
where $\nu$ is the CFL number and $s$ is the maximum wave speed seen on a grid level.

\subsection{Clawpack Implementation}
The algorithm described above is implemented in Clawpack \cite{clawpack}. 
In this subsection, we briefly introduce the implementation of the AMR algorithm in Clawpack in order to better explain the changes added to the code in this study later.

Each grid patch (except those on the coarsest level) has two buffers.
The first buffer (hereafter referred to as \texttt{coarse cell buffer}) of a level $L$ grid patch stores old (at time $t_n$) coarser (level $L-1$) cell values need to be used in computing $C1$.
The second buffer (hereafter referred to as \texttt{conservation fix buffer} accumulates modification terms in \cref{eq:c123}, which are to be added to coarser (level $L-1$) cells in Step \ref*{fig:amr_flow_chart}.7.

Each grid patch (except those on the finest level) also has a lookup table. 
Each table stores information of cells in this grid patch that border nested finer grid patches, including space indices of the cell, which finer grid patch's \texttt{conservation fix buffer} stores modification terms that should be added to this cell in the conservation fix step and memory location of the modification term in that buffer.
When Step \ref*{fig:amr_flow_chart}.7 is conducted between level $L$ and $L-1$, the lookup tables of all grid patches on level $L-1$ are used to find modification terms from the \texttt{conservation fix buffers} of level $L$ grid patches and add those terms to cells in level $L-1$ grid patches.

\begin{figure}[!t]
\centering
\includegraphics[width=2.0in]{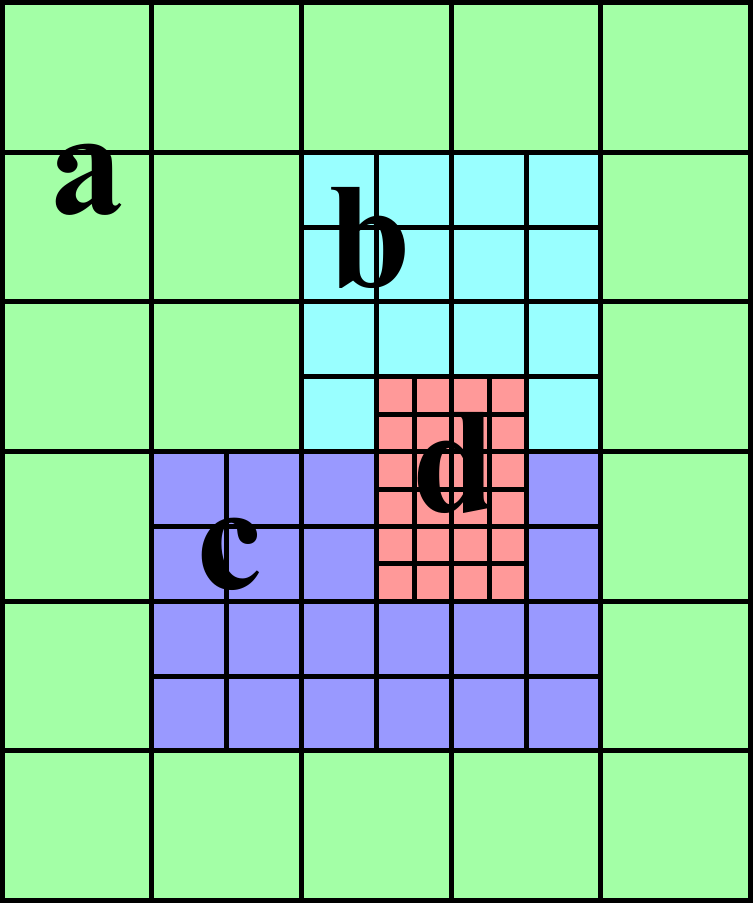}
\caption{Grid patches on level $L-1$, $L$ and $L+1$ (ghost cells not shown). Grid patch $a$ is on level $L-1$, $b$ and $c$ are on level $L$, and $d$ is on level $L+1$.}
\label{fig:grid_3_levels}
\end{figure}

\begin{figure}[!t]
\centering
\includegraphics[width=0.65\linewidth]{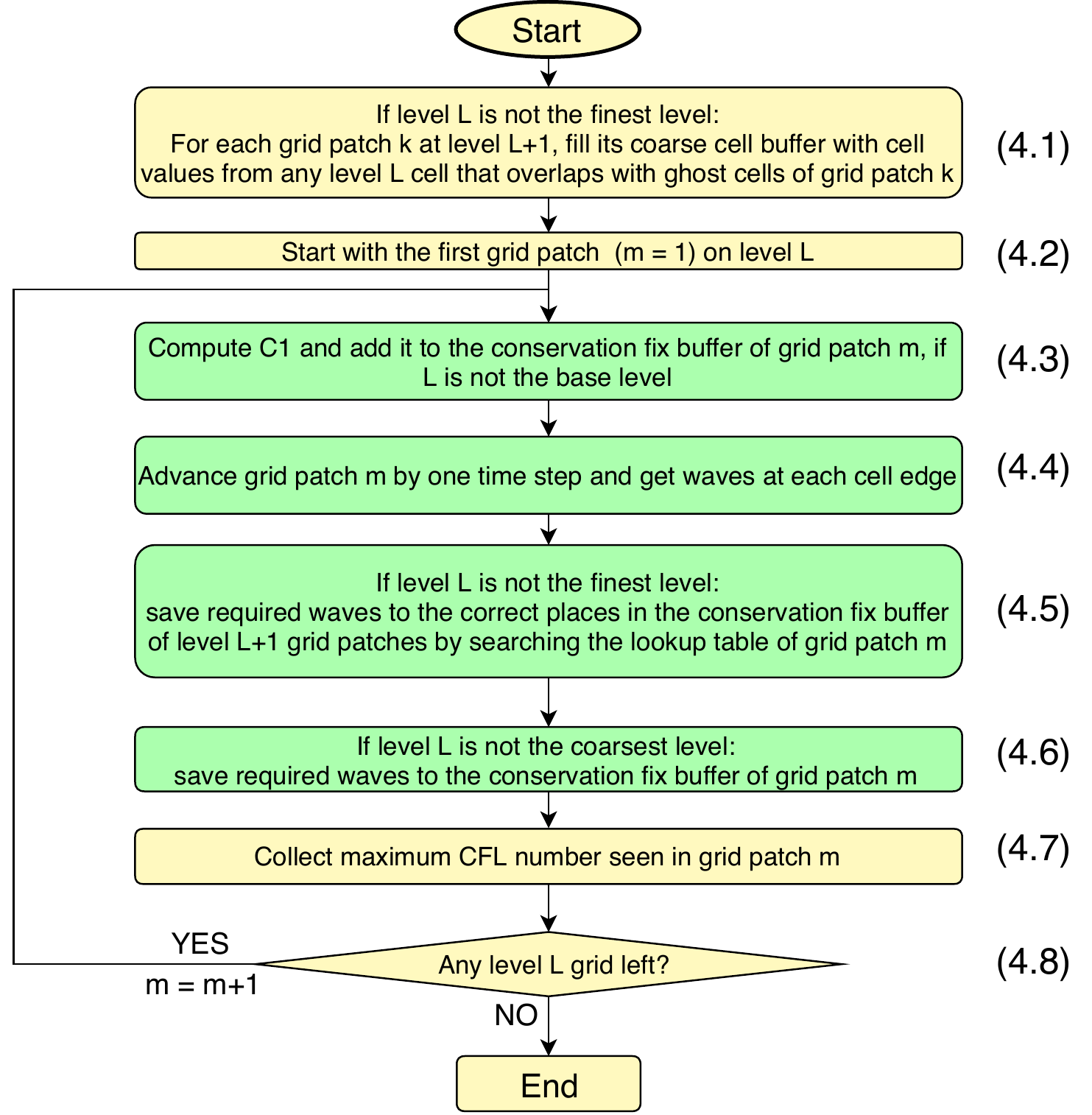}
\caption{Operations included in Step \ref*{fig:amr_flow_chart}.3 of \cref{fig:amr_flow_chart} with respect to level $L$ grid patches.
The steps shown in green are done on the GPU in the algorithm described in \cref{sec:transfer_less}}.
\label{fig:advance_flow_chart}
\end{figure}

Step \ref*{fig:amr_flow_chart}.3 in \cref{fig:amr_flow_chart} includes many sub-steps in addition to advancing the solution in time.
\Cref{fig:advance_flow_chart} is a flow chart that summarizes these operations for level $L$.
We use grid patches on three consecutive levels in \cref{fig:grid_3_levels} to illustrate this flow chart when it is conducted for level $L$ grid patches.
The first step (called \texttt{saving coarse value}) is to save cell values in level $L$ grid patches (grid patches $b$ and $c$ in this case) cells that surround level $L+1$ grid patches (grid patch $d$ in this case) to the \texttt{coarse cell buffers} of level $L+1$ grid patches (grid patch $d$ in this case).
Then the code iterates over all grid patches on level $L$ and does the following.
Starting with grid patch $b$, \cref{eq:c1_1} is computed by solving Riemann problems between pairs of left and right states.
The left (or right) states are old cell values from the $6$ grid patch $a$ cells that surround grid patch $b$, which were stored in the \texttt{coarse cell buffer} of grid patch $b$ when the first step (\texttt{saving coarse value}) was conducted earlier, when level $L-1$ was advanced.
The right (or left) states are cell values in grid patch $b$ that border grid patch $a$ at the current time step and are available.
The resulting waves from solving the Riemann problems are then added to the \texttt{conservation fix buffer} of grid patch $b$.
After computing \cref{eq:c1_1}, the solution in grid patch $b$ is advanced, during which waves $\apdq$ and $\amdq$ are computed at every cell edge.
Then some of these waves are stored into two buffers: 
\begin{enumerate}
    \item Step A: Since level $L$ is not the finest level, waves that emit from the interface between grid patch $b$ and level $L+1$ grid patches (grid patch $d$ in this case) and propagate into the latter are added to the \texttt{conservation fix buffer} of the latter (grid patch $d$ in this case). Note that this is for the conservation fix between level $L+1$ and $L$.
    \item Step B: Since level $L$ is not the coarsest level, waves that emit from the interface between grid patch $b$ and level $L-1$ grid patches (grid patch $a$ in this case) and travel into the latter (\cref{eq:c2_1,eq:c3_1}) are added to the \texttt{conservation fix buffer} of grid patch $b$. Note that this is for the conservation fix between level $L$ and $L-1$.
\end{enumerate}
If Step B above is the second time this step is performed for grid patch $b$, at this time grid patch $b$ already has all modification terms $C1$, $C2$ and $C3$ in its \texttt{conservation fix buffer} since \cref{eq:c2_3,eq:c3_3} were saved to the buffer when the Step A above was conducted for level $L-1$ grid patches (grid patch $a$ in this case) earlier.
In the next step, the maximum wave speed seen in grid patch $b$ while advancing the solution at this time step are collected to update a global maximum CFL number seen on level $L$, which is used to guide adjustment of the time step based on the Courant number limit.
The loop in \cref{fig:advance_flow_chart} now finishes one iteration and the same operations are conducted on the next grid patch on this level (grid patch $c$ in this case).

To guide the code to select an appropriate time step size $\Delta t$ in each time step, a desired CFL number is usually specified by the user before a simulation.
During the simulation, after all grid patches on a level are advanced, the maximum wave speed seen in the process of solving the Riemann problems at every cell edge is collected and used to compute a new time step size for next time step:
\begin{equation}
    \Delta t = \frac{\nu_{desired}\Delta x}{|s|}
    \label{eq:cfl}
\end{equation}
where $\nu_{desired}$ is the specified desired CFL number and $s$ is the maximum wave speed seen.

\section{Code Design and Implementation} \label{sec:design}
In this study, we accelerate the Clawpack package with the Graphics Processing Unit (GPU) by adding new routines and re-writing some of the existing code to expose parallelism in solving hyperbolic PDEs on adaptive meshes.
\subsection{General Design}
There exist many programming models for GPU programming such as CUDA, OpenCL, OpenACC and OpenMP.
In the current study, NVIDIA's CUDA programming model (CUDA 8.0) is used and the implementations are tested on machines where one or more GPUs are attached to a CPU, but with separate physical memory.
A function that is written to run on a GPU is called a GPU kernel.
A GPU kernel must be launched by a CPU thread to be executed on a GPU.
During the launching process, the CPU thread copies all arguments of the function (kernel) from CPU memory to GPU memory, sets up kernel configurations and pushes the execution of the function into the GPU's task queue.
These operations make up the overhead of CPU thread launching a GPU kernel, which can take $10$-$30$ $\mu s$ and can't be neglected when there are many tiny kernels that take only several $\mu s$ to execute.
For example, if each kernel finishes in $5$ $\mu s$ on the GPU but it takes the CPU thread $10$ $\mu s$ to put it into the GPU's task queue, the queue will be empty more than half of the time, during which the GPU is just waiting for new tasks pushed into the queue. 

Clawpack is a tool designed to solve many different hyperbolic systems with the same general high-resolution finite volume method.
For a specific system, a corresponding Riemann solver routine has to be provided.
Other tasks like AMR grid patch construction and destruction, time step size estimation, ghost cells preparation, memory management, synchronization between different levels of grid patches (e.g. Step \ref*{fig:amr_flow_chart}.2, \ref*{fig:amr_flow_chart}.6 and \ref*{fig:amr_flow_chart}.7 of \cref{fig:amr_flow_chart}) are all handled by Clawpack.

In this study, some operations in the AMR algorithm are chosen to be written as GPU kernels executing on the GPU while the others are kept on the CPU, based on their characteristics and data management requirements.
New routines running on the CPU are also added to facilitate memory management, reduce memory footprint, and manage GPU kernels that run asynchronously with respect to the CPU threads.

\subsubsection{A Simple Idea}

The simplest idea of using the GPU to solve a hyperbolic systems with AMR would be to replace the existing part of the CPU code that advances the solution on a single grid patch with a GPU kernel (or multiple GPU kernels) that does the same job, while keeping all other parts of the code unchanged.
Also each time before the solution on a grid patch is advanced, the cell values on the grid patch must be transferred to the GPU memory.
After the solution is advanced, not only the new cell values on the entire grid patch must be transferred back to the CPU memory, but also the left-, right-, up- and down-going waves (for 2D problems) on the entire grid patch, which in total are $5$ times larger than the cell values on the grid patch in terms of memory size.
The wave data on entire grid patches must be transferred back to the CPU because the routines that compute $C1$, $C2$ and $C3$ run on the CPU in this approach. 
Although the time for such data movements can be partially hidden (depending on how long the data movement takes) by overlapping them with kernel execution, the amount of data that must be transferred back to CPU memory turns out to be so large that this time can barely be hidden in many cases.

\subsubsection{A Fix that Transfers Less Data} \label{sec:transfer_less}
A simple fix to this is to compute $C1$, $C2$ and $C3$ on the GPU as well and save these terms to the right place in the buffer of grid patches for conservation fix. 
In this way, the wave data on the entire grid patches does not have to be transferred back to the CPU memory, so the amount of data transferred is reduced by $80\%$.
The four steps colored in pale green in \cref{fig:advance_flow_chart} are the operations that must be done on the GPU to implement this fix.
In order to save the computed $C1$, $C2$ and $C3$ to the right place in memory, the lookup table associated with each grid patch must be in the GPU memory because the GPU kernels now need to access the table on the GPU.
Transferring the lookup tables to the GPU memory turns out to be a cheap operation since 
1) the size of each table is only $O(C)$ where C is the length (measured in number of cells) of all coarse-fine interfaces inside the grid patch that the table is associated with; 2) the tables only need to be transferred to the GPU memory every time regridding takes places on that grid level.

\Cref{fig:advance_flow_chart} shows that for each grid patch on a level, at least one GPU kernel needs to be launched to execute each of Steps \ref*{fig:advance_flow_chart}.3--\ref*{fig:advance_flow_chart}.6 in \cref{fig:advance_flow_chart}.
This can cause many small GPU kernels being executed during the simulation because Steps \ref*{fig:advance_flow_chart}.3, \ref*{fig:advance_flow_chart}.5 and \ref*{fig:advance_flow_chart}.6 for a grid patch usually have much less parallelism than advancing the same grid patch (Step \ref*{fig:advance_flow_chart}.4).
For example, in Step \ref*{fig:advance_flow_chart}.4, Riemann problems need to be solved at each cell edge, the number of which is roughly the same as the number of cells in the grid patch. 
Each GPU thread can be naturally mapped to a cell edge to solve the Riemann problem at the edge.
Hence a GPU kernel that executes Step \ref*{fig:advance_flow_chart}.4 on a grid patch of $a$ by $a$ cells consumes $O(a^2)$ GPU threads.
On the other hand, a GPU kernel that executes Step \ref*{fig:advance_flow_chart}.3 on the same grid patch only consumes $O(a)$ GPU threads since only roughly $4a$ Riemann problems (in the 2D case) need to be solved between its ghost cell values and cell values in its buffer (\cref{eq:c1_1,eq:c1_2}).
Similarly, GPU kernels that execute Steps \ref*{fig:advance_flow_chart}.5 and \ref*{fig:advance_flow_chart}.6 also only consume $O(a)$ GPU threads.
As a result, the overhead of launching kernels dominates the execution of the code and the GPU stays idle for a large portion of time, simply waiting for the CPU thread to push a task into its task queue.

\subsubsection{Merging Small GPU Kernels}
\begin{figure}[!t]
\centering
\includegraphics[width=0.6\linewidth]{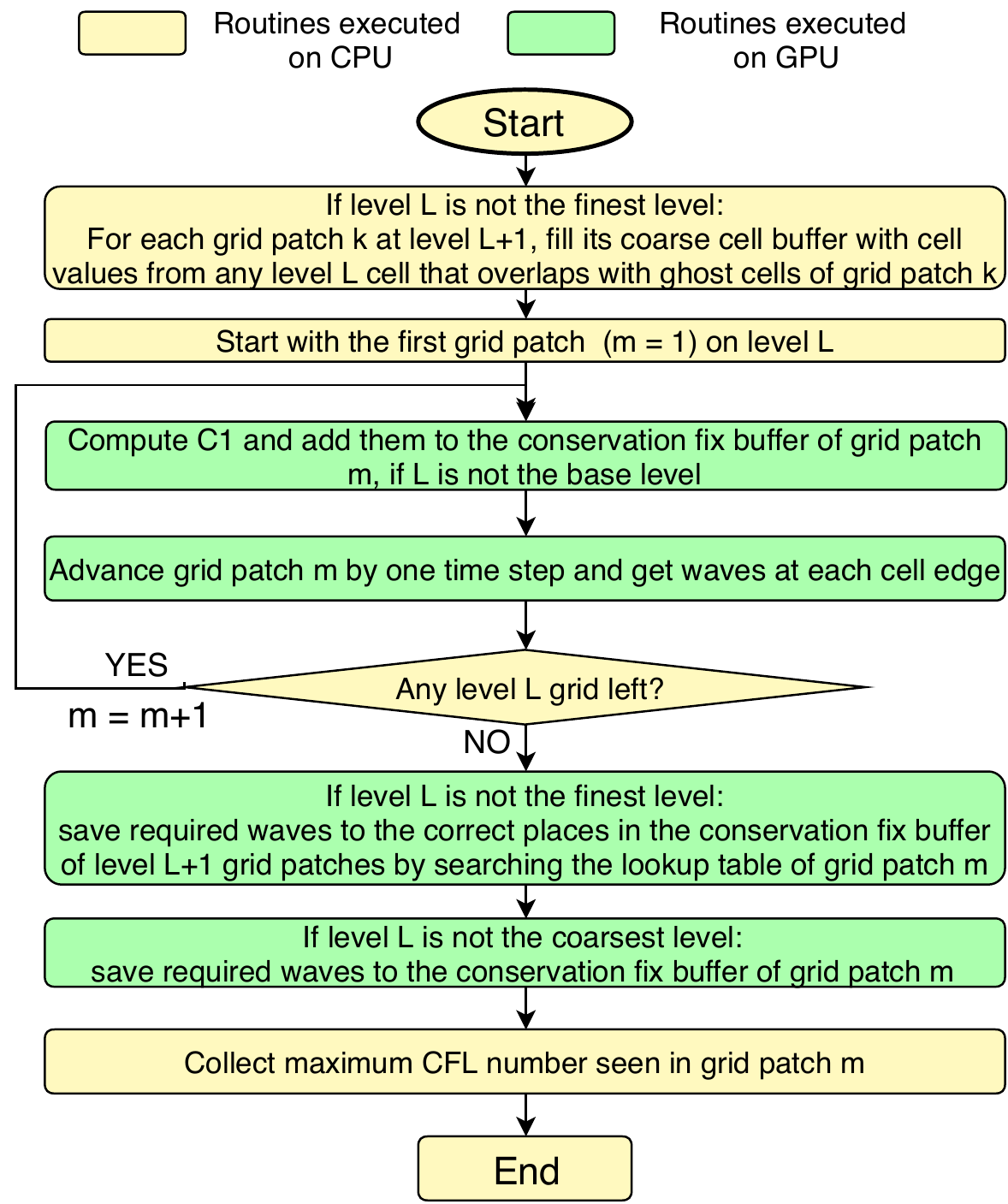}
\caption{Re-design of Step \ref*{fig:amr_flow_chart}.3 in \cref{fig:amr_flow_chart} at level $L$.}
\label{fig:advance_flow_chart_gpu}
\end{figure}

The overhead above can be significantly reduced if we merge some of those small kernels into bigger ones.
\Cref{fig:advance_flow_chart_gpu} shows a new design that conducts the same operations.
Steps \ref*{fig:advance_flow_chart}.5 and \ref*{fig:advance_flow_chart}.6 have been moved out of the loop. 
Only one GPU kernel is launched to perform Step \ref*{fig:advance_flow_chart}.5 for all grid patches on that level.
Similarly, another GPU kernel is launched to perform Step \ref*{fig:advance_flow_chart}.6 regardless of number of grid patches on that level.
One of the challenges is that patches on the same level can have very different sizes.
To deal with this, the two kernels are launched with more threads than what is actually required.
For instance, suppose the largest grid patch on the level has $a$ by $a$ cells (note that $O(a)$ threads are required), the two kernels are then launched with $O(am)$ threads where $m$ is the number of patches on this level.
Although some threads end up with nothing to do this way, especially in the worst case where there is only one huge patch and several small patches at a level, 
some experiments show that this implementation saves a large amount of time compared to the design in \cref{sec:transfer_less}, as the cost of adding more idle threads to the GPU kernel is still much less than total overhead of launching many tiny GPU kernels.

To even further reduce the overhead of launching GPU kernels, the loop in \cref{fig:advance_flow_chart_gpu} iterates from larger grid patches to smaller grid patches.
This is because GPU kernels for smaller grid patches might finish even faster than the overhead of launching the kernel, which makes the GPU idle if future kernels have not been pushed into the task queue.
Launching kernels for the biggest grid patch first gives the system more time to push future GPU tasks into the task queue.

Note that Step \ref*{fig:advance_flow_chart}.3 is still kept in the loop so one GPU kernel is launched for this step for each grid patch.
This is because Step \ref*{fig:advance_flow_chart}.3 for a grid patch must be conducted after cell values on that grid patch have been transferred to the GPU memory.
So if a single merged kernel is used to conduct Step \ref*{fig:advance_flow_chart}.3 for all grids on the level, cell values of all grid patches on this level must be transferred to the GPU memory first, which breaks the data transferring pipeline detailed in \cref{sec:pipeline}.

\subsection{Data Management} \label{sec:data_management}

\subsubsection{The Data Transfer and Kernel Execution Pipeline} \label{sec:pipeline}

Since the GPU and the CPU used in this study have separate physical memory, data has to be transferred through the PCI-e bus between the GPU memory and the CPU memory.
The memory bandwidth of a PCI-e bus is usually very slow (in the order of $\sim 10$ GB/s) compared to the bandwidth of the GPU DRAM memory.
So in principle, data transfer between the GPU memory and the CPU memory should be reduced as much as possible, or the time spent on data transferring should be hidden as much as possible by overlapping data transferring with GPU kernel execution.
If data transferring can be completely hidden, there is no increase in wall time due to moving data between the GPU memory and the CPU memory.

Ideally, if no data has to be transferred between the GPU memory and the CPU memory during the simulation (only at the beginning and near the end of the execution of the code), no extra time needs to be spent on data transferring.
This requires executing all routines including those for the regridding and {\em updating} processes on the GPU.
Some of these processes do not contain as much parallelism as advancing the solution on a grid patch and/or have frequent execution branches, which can hurt the GPU performance significantly.
Avoiding all data transferring also requires keeping data on all levels of grid patches in the GPU memory all the time during the execution, even though only one level of grid patches is advanced at a time.
This study implements a different strategy, which transfers grid patch data between the CPU and GPU memory, can solve a bigger problem since data on only one level of grid patches needs to be in the GPU memory at a time, and leaves routines that the GPU is not good at to be executed by the CPU.
It turns out this strategy introduces almost no extra time on data transferring, even for a problem that contains very few computations relative to the size of data that needs to be transferred (e.g. the acoustic problem in \cref{sec:benchmark}), by using a pipeline detailed in this section to hide data transferring.

\begin{figure}[!t]
\centering
\includegraphics[width=0.7\linewidth]{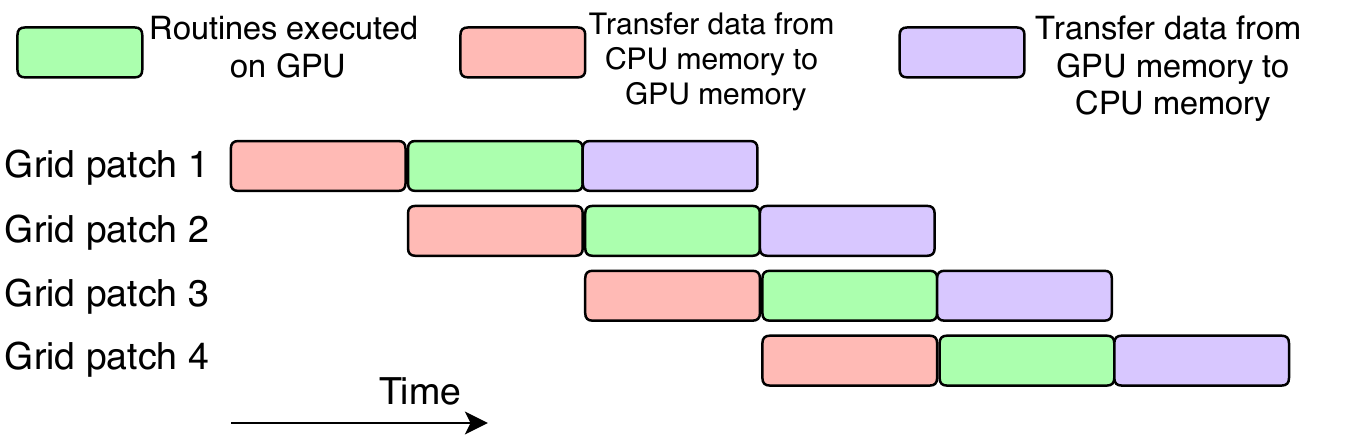}
\caption{A pipeline that hides data transferring in a case where there are four grid patches on the current level.}
\label{fig:data_hiding}
\end{figure}

A simple example of how data transferring can be hidden is shown in \cref{fig:data_hiding}. 
Each rectangular bar represents the time spent on one of the three operations explained in the legend of the figure.
Note that the GPU is idle (does not execute any kernel) only when cell data in Grid patch $1$ is transferred to the GPU memory and when cell data in Grid patch $4$ is transferred to the CPU memory.  
In another word, all other data transferring operations are hidden.

\begin{figure}[!t]
\centering
\includegraphics[width=0.6\linewidth]{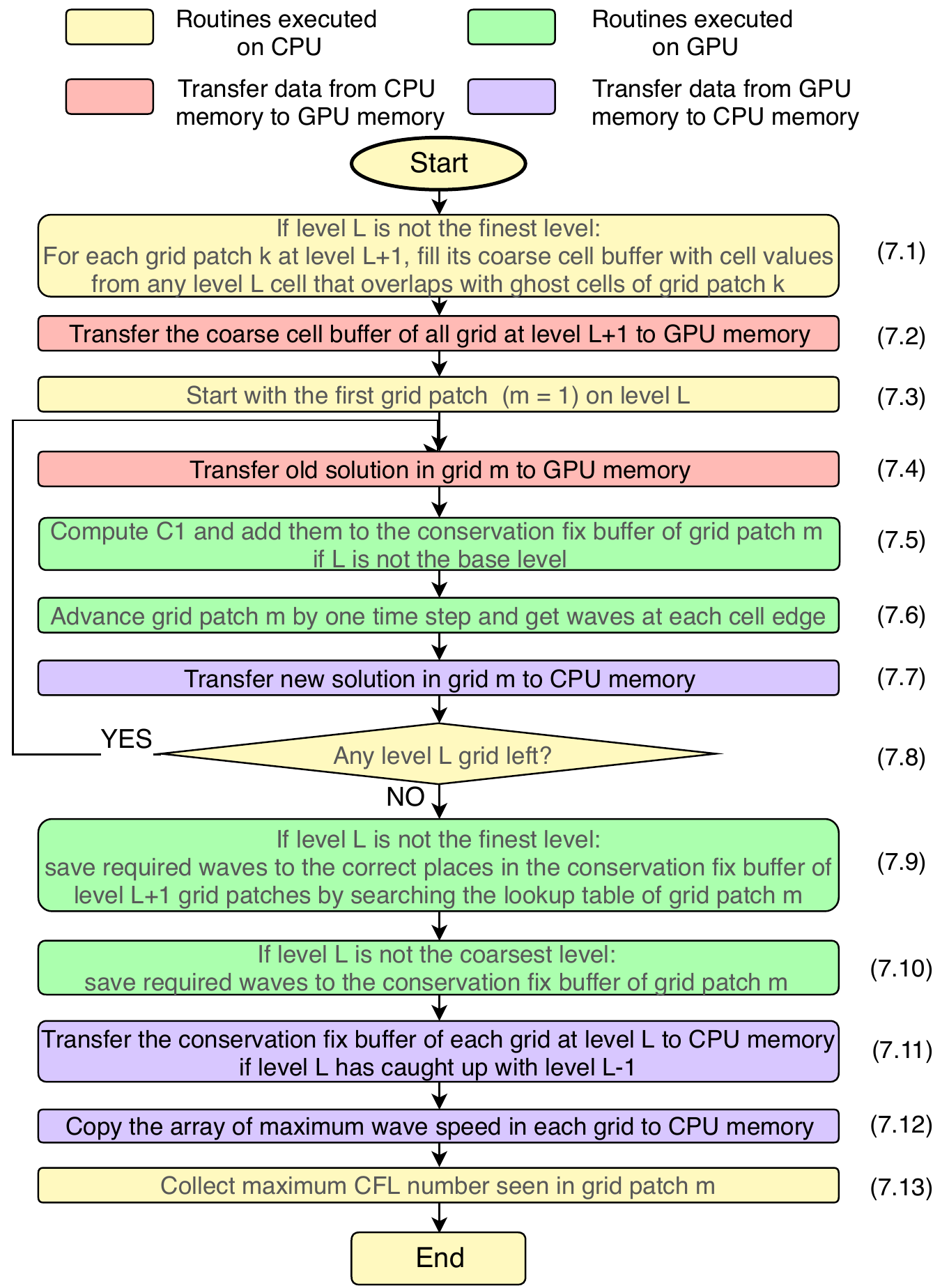}
\caption{Flow chart of Step \ref*{fig:amr_flow_chart}.3 in \cref{fig:amr_flow_chart}, including data transferring between the CPU and the GPU memory.}
\label{fig:data_flow}
\end{figure}

\Cref{fig:data_flow} adds data transferring operations to the flow chart in \cref{fig:advance_flow_chart_gpu}.
Step \ref*{fig:data_flow}.2 of \cref{fig:data_flow} asynchronously transfers the \texttt{coarse cell buffer} of each grid patch on level $L+1$ to the GPU memory.
Note that the flow chart is illustrating operations done when the level $L$ grid patches are to be advanced. 
Although the \texttt{coarse cell buffer} will not be used until the same operations in this flow chart are conducted for level $L+1$, launching the transferring operations here gives more time for the operations to be executed in parallel with computing operations that follows, and thus helps to hide the data transferring.

One of the most useful features in the CUDA programming model is the CUDA streams.
GPU tasks (e.g. GPU kernel execution and data transferring) in different CUDA streams can be executed independently by the GPU if no other restrictions are specified and there are available resources.
GPU tasks in the same CUDA stream must be executed in the order they are launched.
The CPU threads can launch GPU tasks asynchronously, which means the CPU threads can schedule such tasks into CUDA streams, move forward to execute other CPU instructions (including scheduling new GPU tasks) and let the GPU pick what tasks in the CUDA stream to execute in the order restricted by the above rules without interrupting ongoing CPU side instructions.
In the current implementation, each iteration of the loop in \cref{fig:data_flow} is scheduled in a different CUDA stream.
This essentially specifies dependencies of all operations conducted during the entire loop.
Step \ref*{fig:data_flow}.4--\ref*{fig:data_flow}.7 of \cref{fig:data_flow} in a single iteration have dependencies implied by the flow chart: they must be executed in the order specified the arrows in the flow chart.
However, steps in different iterations do not depend on each other and are allow to interleave with each other.
These dependencies allow most of the data transferring (Step \ref*{fig:data_flow}.4 and \ref*{fig:data_flow}.7) to be hidden by kernel execution (Step \ref*{fig:data_flow}.5 and Step \ref*{fig:data_flow}.6).

Usually the tasks from all iterations are scheduled much earlier than when the GPU finishes all tasks.
As a result, a barrier on the CPU side must be inserted right after the loop in \cref{fig:data_flow} since Step \ref*{fig:data_flow}.9 can not be executed until Step \ref*{fig:data_flow}.6 for all grid patches on this level are finished. 

Whenever a grid level catches up in time with the coarser level after a time step, the \texttt{conservation fix buffers} of all grid patches on this level are transferred back to the CPU (Step \ref*{fig:data_flow}.11 of \cref{fig:data_flow}) since they store modification terms $C1$, $C2$ and $C3$ on the GPU side and will be needed in the conservation fix process (Step \ref*{fig:amr_flow_chart}.7).

\subsubsection{Global Reduction for Adapting the Time Step Size}
Each time the solution in each grid patch is advanced on the GPU, a global reduction is conducted to collect the maximum wave speed seen over this grid patch, which is saved to an array in the GPU memory.
After all grid patches of that level are advanced, the array has the maximum wave speed seen in each grid patch and is transferred back to the CPU memory, where another level of reduction is conducted by the CPU to find the maximum wave speed on this grid level.
This maximum wave speed is then used to adjust the time step size, as guided by \cref{eq:cfl}.

\subsubsection{Memory Pool}
Each time a grid patch is advanced or reconstructed from a regridding process, memory blocks must be allocated on the GPU/CPU for storing cell values and waves, and holding buffers for the conservation fix etc., and then freed after the process.
It turns out the overhead of calling the CUDA runtime API to conduct these frequent memory operations can not be neglected, and can even dominate the entire code execution when grid patches are small so the overhead is expensive relative to the time spent on advancing the grid patch.
A user managed memory pool is thus created, which allocates a huge chunk of memory at a time and allocates more chunks when needed. 
All frequent memory allocation and free operations requested by the Clawpack library are through this memory pool, with no need to actually allocate/free system memory.

\subsection{Riemann Solvers} 
For a specific system of hyperbolic equations, normal and transverse Riemann solvers for this system of equations must be provided to advance the solution on grid patches, as described in \cite{leveque2002finite}.
In Clawpack, a routine is called for each row of a grid patch, which sweeps through the 1D slice of the grid, solves Riemann problems to get fluctuations, applies proper limiters, and solves a different Riemann problem for transverse waves.
A similar routine is called for each column of a grid patch.
To perform these operations on the GPU, sweeping of each row in a grid patch is merged into a single GPU kernel to reduce kernel launching overhead.

There can be other dedicated strategies for solving hyperbolic system on the GPU, which carefully design the GPU kernel to map GPU threads to cell interfaces or cells and/or use shared memory to reduce the often-seen memory bottleneck in such a problem.
Since the focus of this study is the on the library level, for the benchmark in the following section a simple and straight-forward implementation that maps each GPU thread to the normal and transverse Riemann problem at one cell interface is used without spending much time on optimizing the performance of the solver itself.

\section{Performance Analysis}\label{sec:benchmark} 
\subsection{Benchmark Setup}
We benchmark the performance of our code by solving a linearized 2D acoustic equation, which can be represented by 
\begin{equation}
    q_t + Aq_x + Bq_y =0
\end{equation}
where
\begin{align}
    q = \begin{bmatrix}
        p \\
        u \\
        v
    \end{bmatrix}, \qquad
    A = \begin{bmatrix}
        0 & K_0 & 0 \\
        \frac{1}{\rho _0} & 0 & 0 \\
        0 & 0 & 0 
    \end{bmatrix}, \qquad
    B = \begin{bmatrix}
        0 & 0 & K_0 \\
        0 & 0 & 0 \\
        \frac{1}{\rho _0} & 0 & 0 
    \end{bmatrix}.
\end{align} 

\begin{figure*}[ht]
\centering

\subfloat[AMR level = $1$]{\includegraphics[width=0.32\textwidth]{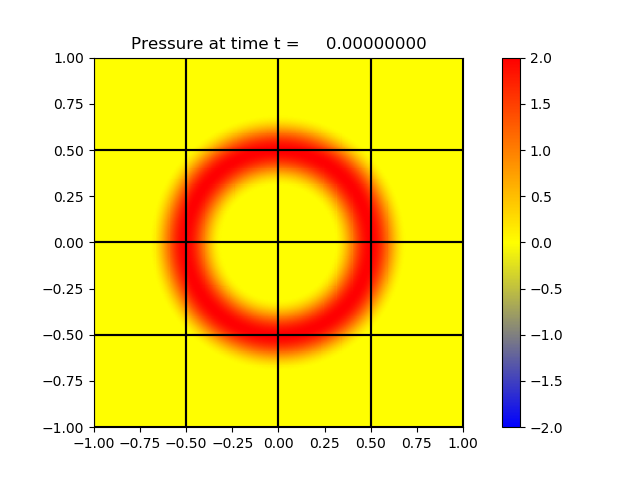} \label{fig:p0}}
\hfill
\subfloat[AMR level = $2$]{\includegraphics[width=0.32\textwidth]{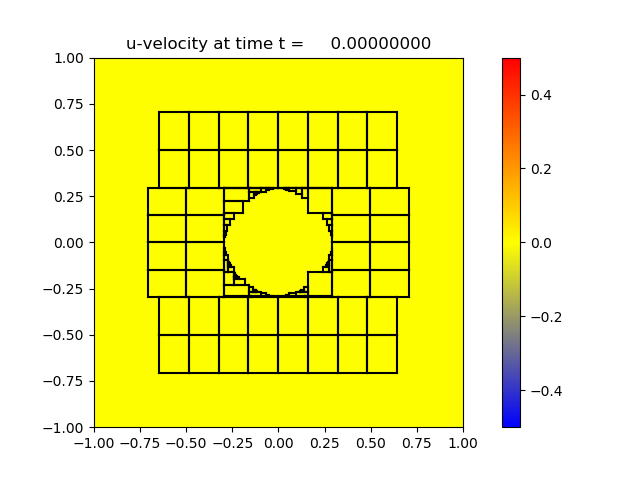} \label{fig:u0}}
\hfill
\subfloat[AMR level = $3$]{\includegraphics[width=0.32\textwidth]{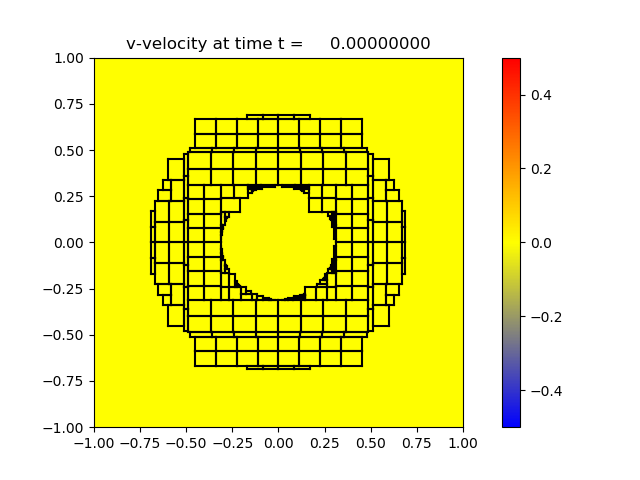} \label{fig:v0}}
\hfill
\subfloat[AMR level = $1$]{\includegraphics[width=0.32\textwidth]{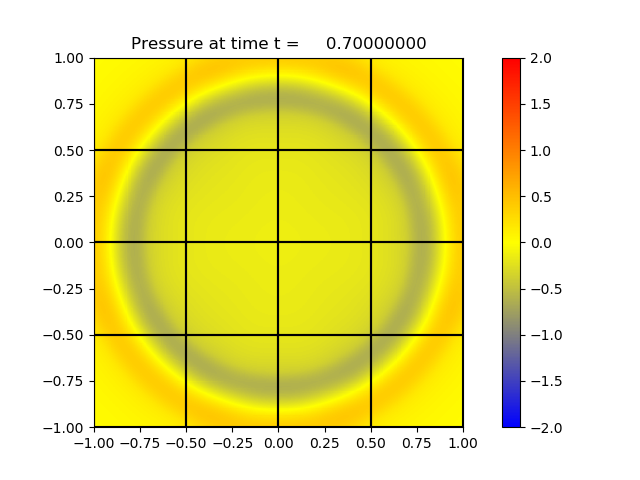} \label{fig:p1}}
\hfill
\subfloat[AMR level = $2$]{\includegraphics[width=0.32\textwidth]{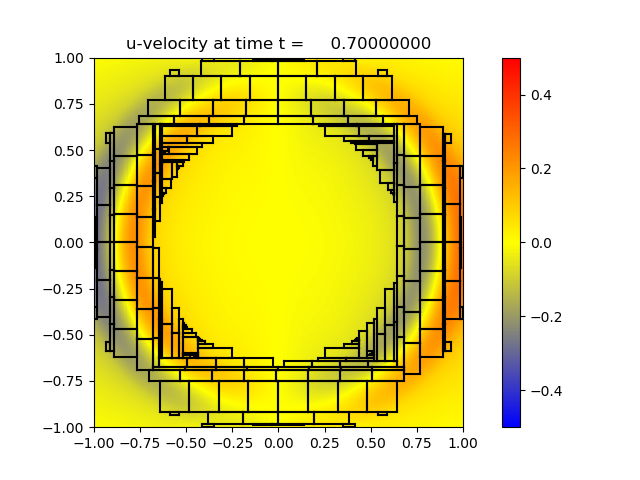} \label{fig:u1}}
\hfill
\subfloat[AMR level = $3$]{\includegraphics[width=0.32\textwidth]{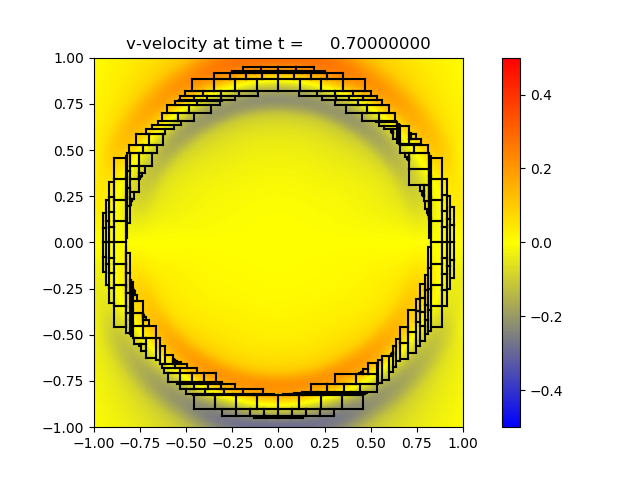} \label{fig:v1}}

\caption{Snapshots of the simulation at $t=0$ (top) and $t=0.7$ (bottom). 
         \protect\subref{fig:p0} and \protect\subref{fig:p1} show pressure field with only level $1$ AMR grid patches;
         \protect\subref{fig:u0} and \protect\subref{fig:u1} show u-velocity field with only level $2$ AMR grid patches;
         \protect\subref{fig:v0} and \protect\subref{fig:v1} show v-velocity field with only level $3$ AMR grid patches.
         Note that only grid patch edges, not grid cells, are showed.
     }
\label{fig:snapshots}
\end{figure*}

In the numerical experiment, an acoustic wave is expanding in the radial direction in a homogeneous medium.
At the beginning of the simulation, a ring-shaped perturbation is added to the pressure field in the domain, as showed in \cref{fig:p0}. 
All four boundaries of the domain have a non-reflecting outflow boundary condition.

A total of 3 levels of AMR grids are used with refinement ratios of 2 between each two levels. 
The computational domain is square with 1000 by 1000 cells on the base level, leading to an effective 4000-by-4000-cell resolution on the finest level.
To ensure grid patch data on an entire grid patch can fit into the L3 cache of the CPU to keep data locality, the size of each grid patch is limited to 260 by 260 in all runs.
A dimensionally unsplit scheme is used incorporate corner transport of waves, as described in \cite{leveque2002finite},
and the Van Leer limiter is applied to the waves.
The problem is simulated for a simulation time of 1 second with approximately 1000 time steps in total.
All floating point numbers are in double precision.
\Cref{fig:snapshots} show several snapshots from the simulation with AMR grid patches.

With the problem and setup described above, we evaluate the performance of the current implementation using four different machines (as many CPU threads as CPU cores are always generated to parallelize the CPU part of the code): 
\begin{enumerate}
    \item a single NVIDIA Kepler K20x GPU with a 16-core AMD Opteron 6274 CPU running at 2.2 GHz as the host;
    \item a single NVIDIA Pascal 100 GPU with a 20-core Intel E5-2698 CPU running at 2.2 GHz as the host (but only 16 CPU threads are used for fair comparison with others);
    \item a single 16-core AMD Opteron 6274 CPU running at 2.2 GHz;
    \item a single 16-core Intel Xeon E-2650 CPU running at 2.0 GHz;
\end{enumerate}
As showed in previous sections, the Clawpack code, after the modification done in this study, consists of jobs that must be done on the host (CPU) and jobs that are done on the device (GPU). 
Hereafter, we will refer to this entire code as the GPU implementation.
The original Clawpack code is referred to as the original CPU code.
The GPU implementation solves the benchmark problem on machine 1) and 2) while the original CPU code that we want to compare with solves the same benchmark problem on machine 3) and 4).

\subsection{Performance Improvement}
\begin{figure*}[ht]
\centering
\subfloat[Comparison between machine 1) and 3)]{\includegraphics[width=0.45\textwidth]{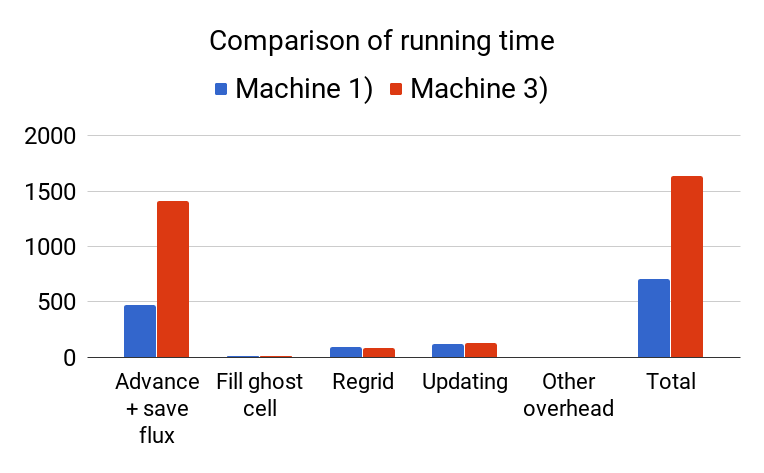} }
\hfil
\subfloat[Comparison between machine 2) and 4)]{\includegraphics[width=0.45\textwidth]{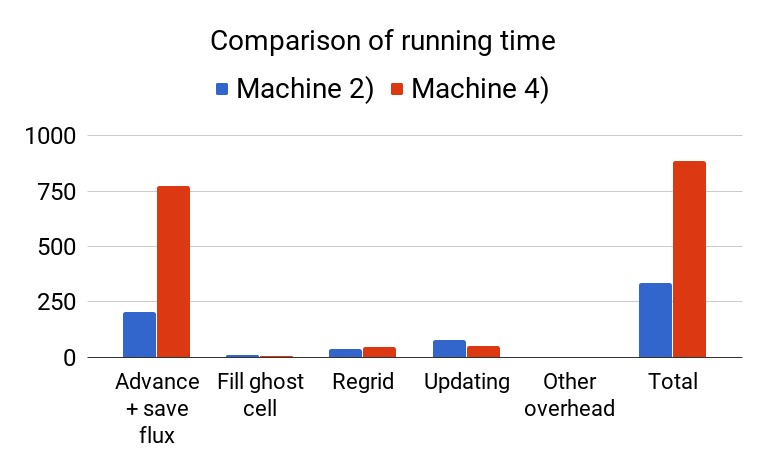} }
\caption{Running time of different sections of the code. From left to right: 1) time on advancing the solution and saving fluxes for later use in conservation fix; 2) time on filling ghost cells; 3) time on regridding; 4) time on the {\em updating} process explained in \cref{sec:amr}; 5) time on all other overhead; 6) total running time.}
\label{fig:time_details}
\end{figure*}

\Cref{fig:time_details} shows running time of different sections of the code as well as total running time on all four machines. 
The cases are setup with \texttt{cutoff} = $0.7$ and \texttt{regrid interval} = 8, which is explained and discussed later in \cref{sec:parameters}.
Note that machine 1) is compared to 3) and machine 2) is compared to 4) such that each pair contains one machine with only CPU running the original CPU code and the other machine consists of the same or similar CPU as the host with an extra GPU.

The comparison shows that most of the execution time is spent on advancing the solution and saving fluxes for later use in the conservation fix, which is completely done on the GPU in the GPU implementation, and exhibits speed up by by a factor of $3.0$ and $3.8$ in the two comparisons.
Most of the other sections of the GPU implementation code are done on the CPU and have similar performance for the corresponding part in the original CPU code.
As a result, using the GPU implementation, the decrease in total running time mostly comes from reduction of time spent advancing the solution and saving fluxes.
But since the other sections are not accelerated by the GPU and still take up a fraction of the total running time, the speed-up of the total running time of the code is observed to be $2.3$ and $2.7$ in the two comparisons.

\subsection{The Influence of AMR Parameters} \label{sec:parameters}
It is best to evaluate the performance of the GPU implementation after some discussion of two parameters, \texttt{cutoff} and \texttt{regrid interval}, that turn out to affect performance significantly on all machines mentioned above.
The parameter \texttt{cutoff} controls how cells flagged to be refined should be grouped together to form grid patches during the regridding process.
It specifies the minimum ratio of the number of flagged cells to all cells on a grid patch to be generated.
An extreme example would be to generate one grid patch for each individual cell, which gives $1.0$ for such a ratio.
In general, a larger \texttt{cutoff} results in a larger collection of smaller grid patches on each level, chosen such that each grid patch does not contain too many unflagged cells (using the algorithm of \cite{BergerRigoutsos}.).
As a result, there are fewer cells in total, but more patches.  This leads to additional overhead, particularly in the GPU version as discussed further below, and so it is important to tune this parameter properly.

The parameter \texttt{regrid interval} specifies how many time steps a level should be advanced before a regridding occurs based on this level.
A larger \texttt{regrid interval} results in fewer regridding operations during the entire simulation, which reduces time spent on the regridding process.
However, if \texttt{regrid interval} is $a$, usually an extra layer of $a$ cells surrounding the original flagged cells will be flagged during the regridding process, such that the waves in the solution do not propagate beyond the refined region before the next regridding process.
As a result, each grid patch is approximately $a$ cells wider at each side and there are more cells in the entire domain. 
Hence, although a larger \texttt{regrid interval} can save time on regridding, it might increase the time spent advancing the solution since there are more cells in total.

\begin{figure*}[ht]
\centering
\subfloat[Total Number of Cells Advanced on All Levels During the Entire Simulation.]{\includegraphics[width=0.48\textwidth]{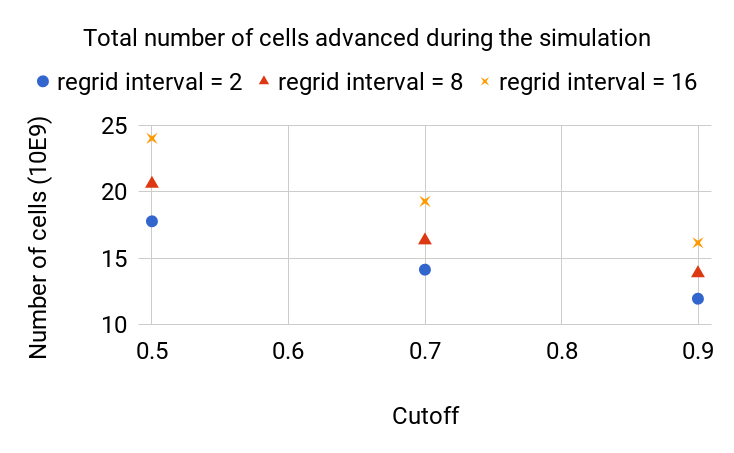} \label{fig:number_of_cells}}
\hfil
\subfloat[Average Number of Grid Patches on All Levels at Each Time Step.]{\includegraphics[width=0.48\textwidth]{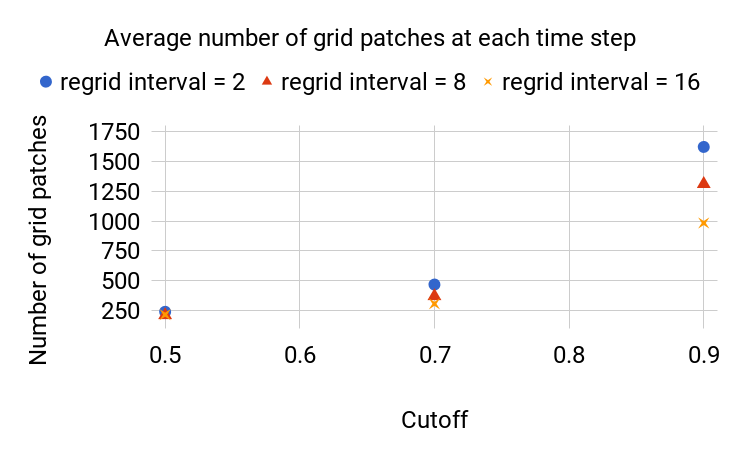} \label{fig:number_of_grids}}
\caption{Number of cells and grid patches for different \texttt{regrid interval} and \texttt{cutoff}.}
\label{fig:compare_cutoff_cells_grids}
\end{figure*}

The effects discussed above can be seen in \cref{fig:number_of_cells,fig:number_of_grids}, which show how the total number of cells and the average number of grid patches at each time step changes as these two parameters are varied.
One can see from the figures that the total number of cells decreases and the average number of grid patches at each time step increases when \texttt{cutoff} increases.
It is interesting to observe that the influence of increasing \texttt{regrid interval} on the total number of cells is similar for different values of \texttt{cutoff}, while the influence of increasing \texttt{regrid interval} on the average number of grid patches at each time step is much smaller when \texttt{cutoff} is small.

\begin{figure*}[ht]
\centering
\subfloat[Machine 1)]{\includegraphics[width=0.45\textwidth]{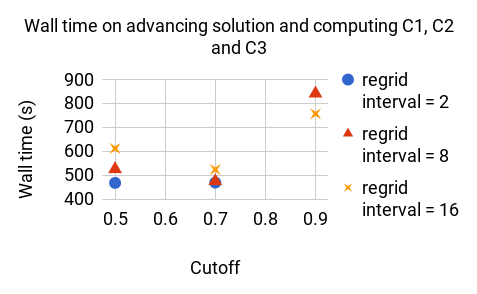} \label{fig:compare_cutoff_advance_time_k20}}
\hfil
\subfloat[Machine 2)]{\includegraphics[width=0.45\textwidth]{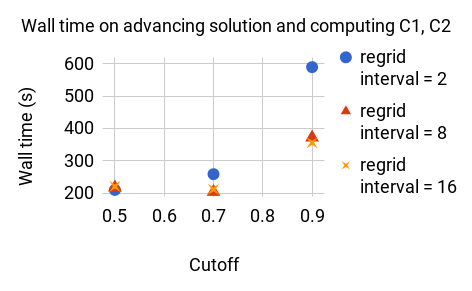} \label{fig:compare_cutoff_advance_time_p100}}
\hfil
\subfloat[Machine 3)]{\includegraphics[width=0.45\textwidth]{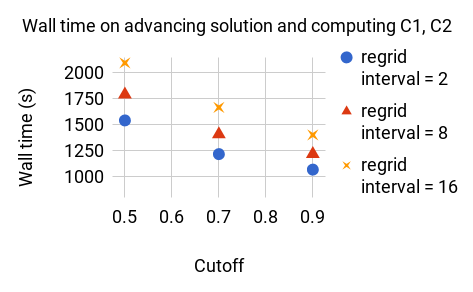} \label{fig:compare_cutoff_advance_time_amd}}
\hfil
\subfloat[Machine 4)]{\includegraphics[width=0.45\textwidth]{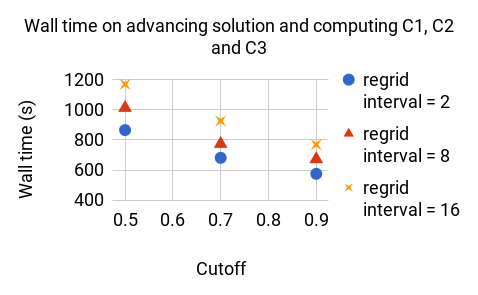} \label{fig:compare_cutoff_advance_time_xeon}}
\caption{Wall time on advancing the solution and computing C1, C2 and C3}
\label{fig:compare_cutoff_advance_time}
\end{figure*}

\Cref{fig:compare_cutoff_advance_time} shows the time spent advancing the solution and computing the $C1$, $C2$ and $C3$ terms. 
Comparing the trend in \cref{fig:compare_cutoff_advance_time_amd,fig:compare_cutoff_advance_time_xeon} with \cref{fig:number_of_cells,fig:number_of_grids} clearly shows that the time spent on this part in the original CPU code is mainly affected by the total number of cells, not the average number of grid patches.
On the other hand, the trend in \cref{fig:compare_cutoff_advance_time_k20,fig:compare_cutoff_advance_time_p100} indicates that the time spent on this part in the GPU implementation is more affected by the average number of grid patches than by the total number of cells.
This is because the overhead of launching CUDA kernels to advance every grid patch is high relative to the execution of the kernel itself, sometimes even longer than the latter if the grid patch is very small. 
Such overhead almost does not depend on the total number of cells but is approximately in proportion to the number of grid patches.
As a result, the overhead of launching CUDA kernels becomes dominant when there are a large fraction of small grid patches.
This indicates that increasing the total number of cells reasonably (having more ``unnecessary'' cells by choosing small \texttt{cutoff} and larger \texttt{regrid interval}) to reduce the number of small patches, as well as reducing the regridding frequency, is an effective strategy for the GPU implementation.
On the other hand, since the overhead of calling CPU functions to advance a grid patch is very low compared to the execution of the function itself for the original CPU code, the total number of cells that need to be advanced is the main resource that affects this running time for the original CPU code on machine 3) and machine 4), rather than number of grid patches.

It is easy to draw the conclusion from \cref{fig:compare_cutoff_advance_time} that for the original CPU code, a smaller \texttt{regrid interval} and larger \texttt{cutoff} is preferable.
For the GPU implementation, a different combination of the two parameters should be chosen.
It can be seen that for a fixed \texttt{regrid interval}, although increasing \texttt{cutoff} reduced the total number of cells needing to be advanced, the time spent advancing the solution on all grids (and computing $C1$, $C2$ and $C3$) increases dramatically when \texttt{cutoff} increases beyond $0.7$ for any \texttt{regrid interval} because many small grid patches are generated for a \texttt{cutoff} of $0.9$.
For a small \texttt{cutoff}, different \texttt{regrid interval} give similar running time.

\begin{figure*}[ht]
\centering
\subfloat[Machine 1)]{\includegraphics[width=0.45\textwidth]{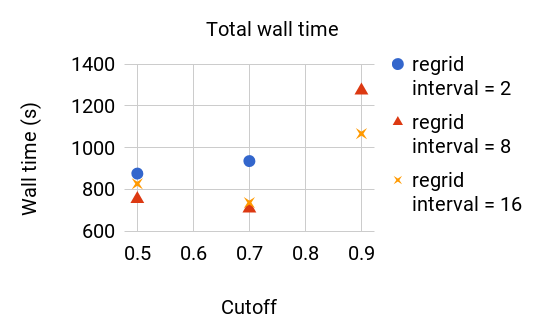} }
\hfil
\subfloat[Machine 2)]{\includegraphics[width=0.45\textwidth]{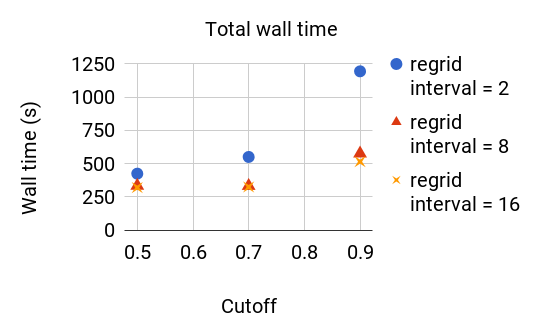} }
\hfil
\subfloat[Machine 3)]{\includegraphics[width=0.45\textwidth]{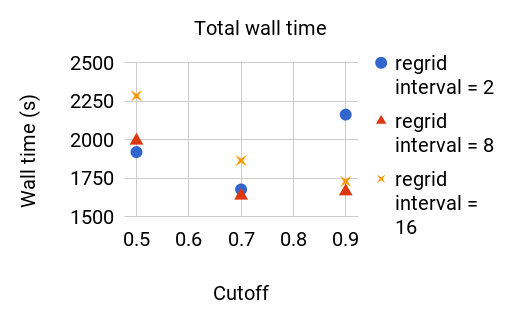} } 
\hfil
\subfloat[Machine 4)]{\includegraphics[width=0.45\textwidth]{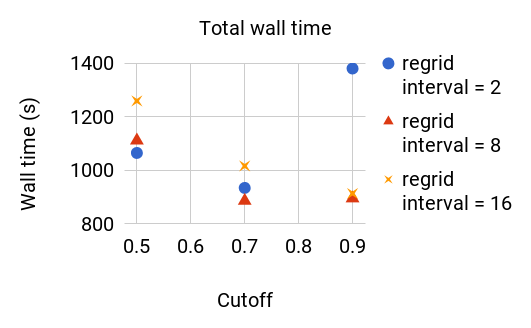} }
\caption{Total wall time. (one data point for regrid interval = 2 on machine 1) is missing because the GPU run out of memory.)}
\label{fig:compare_cutoff_total_time}
\end{figure*}

\Cref{fig:compare_cutoff_total_time} shows that the two parameters have a significant impact on the total running time for both the GPU implementation and the original CPU code.
Here the total running time includes time spent regridding and updating cell values from fine level to coarse level (the {\em updating} process) in addition to the time spent advancing the solution discussed above.
This is why although a combination of \texttt{cutoff} = $0.9$ and \texttt{regrid interval} = $2$ gives the shortest time spent advancing the solution for the original CPU code, this is not the best combination to get the shortest total running time.
Using a \texttt{regrid interval} of $2$, the code spends almost $8$ times longer on regridding than when using a \texttt{regrid interval} of $16$.
For the GPU implementation, the total running time is much more sensitive to an inappropriate choice of these two parameters. 
This can be seen from the fact that, with the worst choice of the two parameters, the original CPU code runs only $1.5$ times slower than the case where the two parameters are chosen to give the best performance (among the nine combinations studied here).
The factor is as large as $4$ for the GPU implementation, and it can be seen that the GPU implementation with the worst parameter setup (\texttt{cutoff} = $0.9$, \texttt{regrid interval} = $2$) runs much slower than any other case.
This worst case specifies very thin buffer layers ($2$-cell wide) around flagged cells so fewer cells are flagged before they are grouped into grid patches, and many large grid patches are cut into very small grid patches to satisfy the requirement from \texttt{cutoff}.
\Cref{fig:compare_cutoff_total_time} suggests that a combination of \texttt{cutoff} = $0.7$ and \texttt{regrid interval} = $8$ seems to be the optimum choice for the original CPU code to get the best performance for this benchmark while for the GPU implementation, a wider range can be chosen to get the shortest running time: any combination from \texttt{cutoff} = $0.5, 0.7$ and \texttt{regrid interval} = $8, 16$ usually gives performance similar to the best.
Note that this ``sweet spot'' might change when the code solves a different problem with much more computational work per cell.  Once the cost of advancing some extra cells becomes comparable to the cost of regridding, one might be able to save some time advancing the solution by reducing the total number of cells, at the cost of doing the (relatively cheap) regridding process more frequently (a smaller \texttt{regrid interval}).

Since the total running time is affected by the two parameters studied above, 
in practice it is more reasonable to compare the running time of the GPU implementation and the original CPU code running at their ``sweet spot'' (best combination of \texttt{cutoff} and \texttt{regrid interval}) 
The ``sweet spot'' is \texttt{cutoff} = $0.7$ and \texttt{regrid interval} = $8$ for the original CPU code on both machine 3) and 4).
The GPU implementation has the same ``sweet spot'' on machine 1) while on machine 2), it achieves the best performance with a different choice where \texttt{cutoff} = $0.5$ and \texttt{regrid interval} = $16$.
Using these results, the GPU implementation on machine 1) is $2.1$ times faster than the original CPU code running on machine 3), while the GPU implementation running on machine 2) is $2.7$ times faster than the original CPU code running on machine 4).

\subsection{Challenges Induced by AMR} \label{sec:challenges}
\begin{figure*}[ht]
\centering
\subfloat[Average Wall Time Per Time Step]{\includegraphics[width=0.3\textwidth]{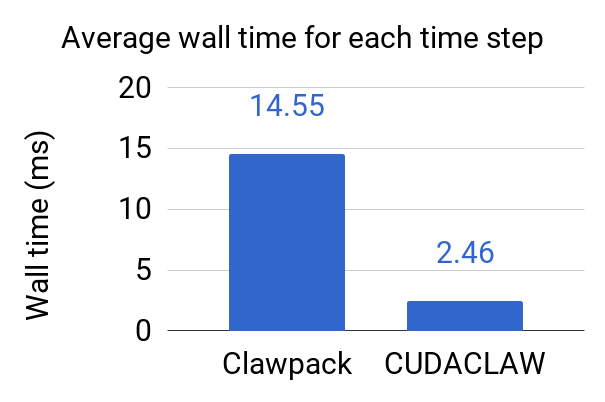} \label{fig:time_each_step}}
\hfil
\subfloat[Load and Write Operations in CUDA Kernels in Clawpack]{\includegraphics[width=0.25\textwidth]{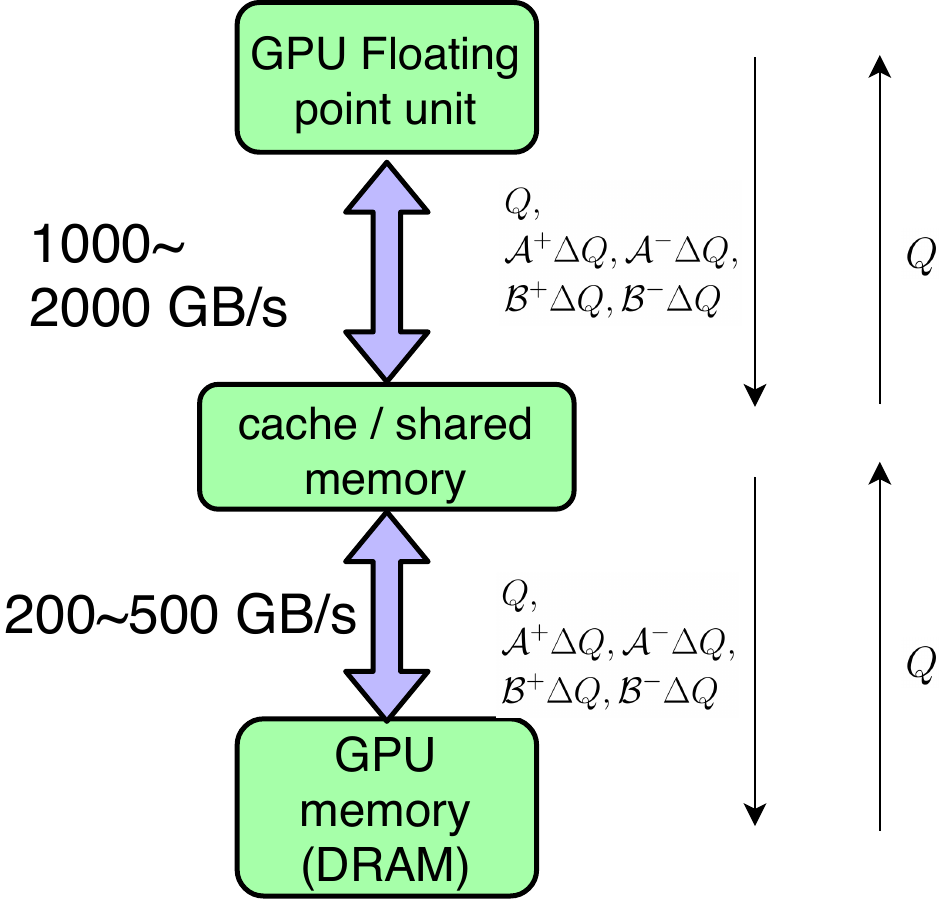} \label{fig:cache_amrclaw}}
\hfil
\subfloat[Load and Write Operations in CUDA Kernels in CUDACLAW]{\includegraphics[width=0.25\textwidth]{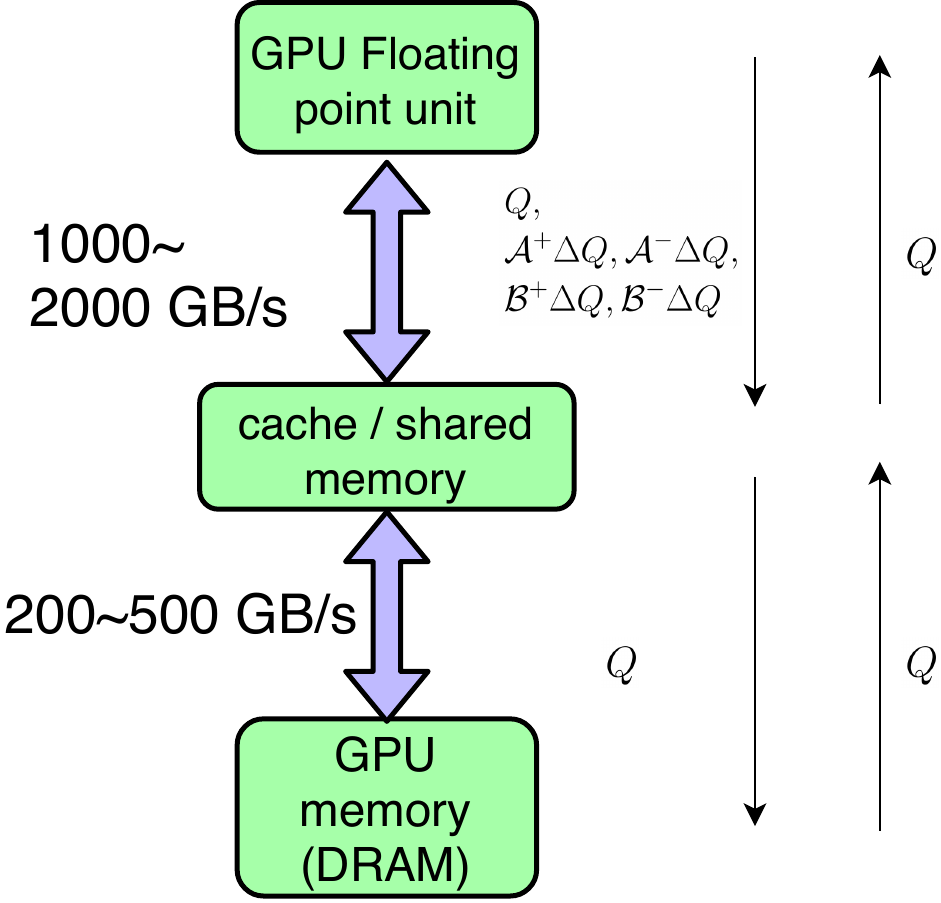} \label{fig:cache_cudaclaw}}
\caption{Comparison of two GPU codes that solve the same problem. The new implementation is designed with AMR ability (Clawpack), while CUDACLAW was not designed to handle adaptive refinement.}
\label{fig:compare_amrclaw_cudaclaw}
\end{figure*}

To reveal the challenges in optimizing the code performance when adaptive refinement is used, another numerical experiment is conducted that solves the same problem but without using AMR.
The GPU implementation of Clawpack and CUDACLAW \cite{ohannessian2018cudaclaw} are compared in the experiment.
In Clawpack, AMR is ``turned off'' by simply setting maximum refine level to $1$.
\cref{fig:time_each_step} shows that the average wall time for each time step advanced in CUDACLAW is about $5$ times shorter than in this new GPU implementation of Clawpack that allows AMR.
This is because the bottleneck for performance in many such problems is memory bandwidth and CUDACLAW consumes much less bandwidth.
\cref{fig:cache_amrclaw,fig:cache_cudaclaw} show data that is loaded from and written to GPU memory (DRAM) in the solver of the two codes.
Since some waves must be selected and stored to buffer for the conservation fix, the GPU implementation of Clawpack must write the $4$ data arrays that store the waves ($\apdq$ etc.) to GPU memory (DRAM), which consumes $4$ times more bandwidth than CUDACLAW. 
The latter is carefully designed to keep all intermediate data in cache and only needs to write the data array for the solution $Q$ to GPU memory (DRAM). 
As a comparison, it is worth noting that other studies also achieve similar speed-up around $2\sim3$ (16-CPU-core-to-1-GPU comparison) when AMR is used, as reviewed in section \ref{sec:introduction}. 
The speed-up can vary depending on whether regridding and updating processes are included in the comparison and on characteristics of the PDEs being solved.

The need for reading/writing additional arrays is largely due to the conservation fix-up applied at each step. While this may be important for some problems, in practice it has been found to have little effect on the solution. For linear problems that are not in conservation form, such as acoustics in a heterogeneous medium, there is no reason to even expect conservation. 
A variant of the code that does not apply this fix is currently under development and preliminary results suggest a further doubling of speed-up is possible.

\section{Conclusion}
This paper describes the design and implementation of a GPU extension to the Clawpack library for solving hyperbolic problems using the wave propagation algorithm with adaptive mesh refinement.
A maximum speed-up of $2.7$ is observed for the benchmark acoustics problem considered here.
The use of adaptive grids requires solving extra Riemann problems and saving the required waves near grid patch edges in order to preserve global conservation.
Doing these operations on the GPU avoids the necessity of transferring all waves back to GPU. 
The sensitivity of the original CPU code and the new GPU implementation to two basic parameters, \texttt{cutoff} and \texttt{regrid interval}, that controls the AMR algorithm are compared and discussed to draw some recommendations for setting the two parameters to achieve better performance.


\appendix
\section*{Acknowledgments}
The first author would like to thank Weiqun Zhang, Max Katz and Ann Almgren for many discussions with them during a summer internship at Lawrence Berkeley National Lab, supported by the AMReX project, which inspired many ideas and design strategies chosen in this work.  This work was also supported in part by NSF grant EAR-1331412 and the University of Washington Department of Applied Mathematics.

\bibliographystyle{siamplain}
\bibliography{amr}

\begin{thebibliography}{10}

\bibitem{beckingsale2015resident}
{\sc D.~Beckingsale, W.~Gaudin, A.~Herdman, and S.~Jarvis}, {\em Resident
  block-structured adaptive mesh refinement on thousands of graphics processing
  units}, in Parallel Processing (ICPP), 2015 44th International Conference on,
  IEEE, 2015, pp.~61--70.

\bibitem{beckingsale2014parallel}
{\sc D.~Beckingsale, W.~Gaudin, R.~Hornung, B.~Gunney, T.~Gamblin, J.~Herdman,
  and S.~Jarvis}, {\em Parallel block structured adaptive mesh refinement on
  graphics processing units}, tech. report, Lawrence Livermore National
  Laboratory (LLNL), Livermore, CA (United States), 2014.

\bibitem{berger1989local}
{\sc M.~J. Berger and P.~Colella}, {\em Local adaptive mesh refinement for
  shock hydrodynamics}, Journal of computational Physics, 82 (1989),
  pp.~64--84.

\bibitem{BergerLeVeque1998}
{\sc M.~J. Berger and R.~J. LeVeque}, {\em Adaptive mesh refinement using
  wave-propagation algorithms for hyperbolic systems}, SIAM J. Numer. Anal., 35
  (1998), pp.~2298--2316, \url{https://doi.org/10.1137/S0036142997315974}.

\bibitem{BergerRigoutsos}
{\sc M.~J. Berger and I.~Rigoutsos}, {\em An algorithm for point clustering and
  grid generation}, IEEE Trans. Sys. Man \& Cyber., 21 (1991), pp.~1278--1286.

\bibitem{bryan2014enzo}
{\sc G.~L. Bryan, M.~L. Norman, B.~W. O'Shea, T.~Abel, J.~H. Wise, M.~J. Turk,
  D.~R. Reynolds, D.~C. Collins, P.~Wang, S.~W. Skillman, et~al.}, {\em Enzo:
  An adaptive mesh refinement code for astrophysics}, The Astrophysical Journal
  Supplement Series, 211 (2014), p.~19.

\bibitem{burstedde2014forestclaw}
{\sc C.~Burstedde, D.~Calhoun, K.~Mandli, and A.~R. Terrel}, {\em Forestclaw:
  Hybrid forest-of-octrees amr for hyperbolic conservation laws}, Parallel
  Computing: Accelerating Computational Science and Engineering (CSE), M.
  Bader, A. Bode, and H.-J. Bungartz, Eds. Amsterdam: IOS Press BV,  (2014),
  pp.~253--262.

\bibitem{burstedde2010extreme}
{\sc C.~Burstedde, O.~Ghattas, M.~Gurnis, T.~Isaac, G.~Stadler, T.~Warburton,
  and L.~Wilcox}, {\em Extreme-scale amr}, in Proceedings of the 2010 ACM/IEEE
  International Conference for High Performance Computing, Networking, Storage
  and Analysis, IEEE Computer Society, 2010, pp.~1--12.

\bibitem{Calhoun2017}
{\sc D.~Calhoun and C.~Burstedde}, {\em {ForestClaw : A parallel algorithm for
  patch-based adaptive mesh refinement on a forest of quadtrees}},
  arXiv:1703.03116,  (2017).

\bibitem{clawpack}
{\sc {Clawpack Development Team}}, {\em Clawpack software}, 2017,
  \url{https://doi.org/10.5281/zenodo.262111}, \url{http://www.clawpack.org}.
\newblock Version 5.4.0.

\bibitem{fryxell2000flash}
{\sc B.~Fryxell, K.~Olson, P.~Ricker, F.~Timmes, M.~Zingale, D.~Lamb,
  P.~MacNeice, R.~Rosner, J.~Truran, and H.~Tufo}, {\em Flash: An adaptive mesh
  hydrodynamics code for modeling astrophysical thermonuclear flashes}, The
  Astrophysical Journal Supplement Series, 131 (2000), p.~273.

\bibitem{leveque2002finite}
{\sc R.~J. LeVeque}, {\em Finite volume methods for hyperbolic problems},
  vol.~31, Cambridge university press, 2002.

\bibitem{mandli2016clawpack}
{\sc K.~T. Mandli, A.~J. Ahmadia, M.~Berger, D.~Calhoun, D.~L. George,
  Y.~Hadjimichael, D.~I. Ketcheson, G.~I. Lemoine, and R.~J. LeVeque}, {\em
  Clawpack: building an open source ecosystem for solving hyperbolic pdes},
  PeerJ Computer Science, 2 (2016), p.~e68.

\bibitem{ohannessian2018cudaclaw}
{\sc H.~G. Ohannessian, G.~Turkiyyah, A.~Ahmadia, and D.~Ketcheson}, {\em
  Cudaclaw: A high-performance programmable gpu framework for the solution of
  hyperbolic pdes}, arXiv preprint arXiv:1805.08846,  (2018).

\bibitem{schive2010gamer}
{\sc H.-Y. Schive, Y.-C. Tsai, and T.~Chiueh}, {\em Gamer: a graphic processing
  unit accelerated adaptive-mesh-refinement code for astrophysics}, The
  Astrophysical Journal Supplement Series, 186 (2010), p.~457.

\bibitem{schive2017gamer}
{\sc H.-Y. Schive, J.~A. ZuHone, N.~J. Goldbaum, M.~J. Turk, M.~Gaspari, and
  C.-Y. Cheng}, {\em Gamer-2: a gpu-accelerated adaptive mesh refinement
  code--accuracy, performance, and scalability}, arXiv preprint
  arXiv:1712.07070,  (2017).

\bibitem{terrel2013manyclaw}
{\sc A.~R. Terrel and K.~T. Mandli}, {\em Manyclaw: Slicing and dicing riemann
  solvers for next generation highly parallel architectures}, arXiv preprint
  arXiv:1308.1464,  (2013).

\bibitem{wang2010adaptive}
{\sc P.~Wang, T.~Abel, and R.~Kaehler}, {\em Adaptive mesh fluid simulations on
  gpu}, New Astronomy, 15 (2010), pp.~581--589.

\end{thebibliography}
\end{document}